\documentclass[preprint]{aa}
\usepackage{graphicx}
\usepackage[breaklinks,colorlinks,citecolor=blue,linkcolor=blue]{hyperref}
\usepackage{txfonts}
\usepackage{epstopdf}

\newcommand{\hctn}{HC$_3$N}
\newcommand{\hcfn}{HC$_5$N}
\newcommand{\hcsn}{HC$_7$N}
\newcommand{\hcnn}{HC$_9$N}
\newcommand{\cts}{C$_3$S}

\newcommand{\cfh}{C$_4$H}
\newcommand{\chtcch}{CH$_3$CCH}
\newcommand{\cfht}{C$_4$H$_2$}

\newcommand{\cctht}{c-C$_3$H$_2$}

\newcommand{\kms}{$\mathrm{km~s^{-1}}$}

\newcommand{\comment}[1]{}
\newcommand{\colhead}[1]{#1}
\newcommand{\nodata}{}

\begin{document}

\title{A particular carbon-chain-producing region:\\
   L1489 starless core}

\author{Yuefang Wu\inst{1}\and
Lianghao Lin\inst{1,2,3}\and
Xunchuan Liu\inst{1}\and
Xi Chen\inst{4,5}\and
Tie Liu\inst{6,7}\and
Chao Zhang\inst{1,8}\and
Binggang Ju\inst{3}\and
Jinghua Yuan\inst{9}\and
Junzhi Wang\inst{4}\and
Zhiqiang Shen\inst{4}\and
Kee-Tae Kim\inst{6}\and
Sheng-Li Qin\inst{8}\and
Juan Li \inst{4}\and
Hongli Liu\inst{10}\and
Tianwei Zhang\inst{1}\and
Ye Xu \inst{3}\and
Qinghui Liu\inst{4}
}

\institute{
        Department of Astronomy, Peking University, 100871, Beijing, China\\ \email{ywu@pku.edu.cn}
        \and
                School of Astronomy and Space Sciences, University of Science and Technology of China, 96 Jinzhai Road, Hefei, 230026, China
        \and
        Purple Mountain Observatory and Key Laboratory of Radio Astronomy, Chinese Academy of Sciences, 8 Yuanhua Road, Nanjing, 210034, China
        \and
                Shanghai Astronomical Observatory, Chinese Academy of Sciences,
                        Shanghai 200030, China
        \and
        Center for Astrophysics, GuangZhou University, Guangzhou, 510006, China
        \and
        Korea Astronomy and Space Science Institute, 776 Daedeokdae-ro,
                        Yuseong-gu, Daejeon 34055, Korea
        \and
        East Asian Observatory, 660 North A'ohoku Place, Hilo, HI 96720, USA;
        \and
        Department of Astronomy, Yunnan University,
                        Kunming, 650091, China
        \and
                National Astronomical Observatories, Chinese Academy of Sciences, 20A Datun Road, Chaoyang District, Beijing 100101, China
        \and
                Department of Physics, The Chinese University of Hong Kong,
                        Shatin, NT, Hong Kong SAR
      }

\abstract{
We detected carbon-chain molecules (CCMs) HC$_{2n+1}$N (n=1-3) and C$_3$S in K$_u$ band as well as {high-energy excitation lines including C$_4$H N=9-8, J=17/2-15/2, 19/2-17/2,  and CH$_3$CCH J=5-4, K=2} in the 3 mm band toward a starless core called the eastern molecular core (EMC) of L1489 IRS. Maps of all the observed lines were also obtained.
Comparisons with a number of early starless cores and WCCC source L1527 show that the column densities of C$_4$H and CH$_3$CCH are close to those of L1527, and the   CH$_3$CCH column densities of the EMC and L1527 are slightly higher than those of TMC-1.
The EMC and L1527 have similar \cts~column densities, but they  are much lower than those of all the starless cores, with only 6.5\% and 10\% of the TMC-1 value, respectively.
The emissions of the N-bearing species of the EMC and L1527 are at the medium level of the starless cores. These comparisons  show that the CCM emissions in the EMC are similar to those of L1527, though L1527 contains a protostar. Although dark and quiescent, the EMC is
warmer and at a later evolutionary stage than classical carbon-chain--producing regions  in the cold, dark, quiescent early phase.
The PACS, SPIRE, and SCUBA maps evidently show that the L1489 IRS seems to be the heating source of the EMC. Although
it is located at the margins of the EMC,  its bolometric luminosity and bolometric temperature are relatively high. 
Above all, the EMC is a rather particular carbon-chain-producing region and is quite significant for CCM science.}

\maketitle


\section{Introduction} \label{sec:intro}
  Carbon-chain molecules (CCMs), including radicals, have the largest number of atoms among interstellar molecules found up until now\footnote{www.astro.uni-koeln.de/cdms/molecules}.
They have a large mass range and have different excitation energies and active states, and as a result they play important roles in interstellar chemical and physical processes.
 Although it is difficult for them to exist in terrestrial conditions, they were detected in star-forming regions and circumstellar envelopes of evolved stars \citep{1971ApJ...163L..35T,1976ApJ...205L.173A,1992ApJ...392..551S,2018ApJ...854..133T,1978A&A....70L..37W,
2006PNAS..10312274Z,2017A&A...606A..74Z}.

 In molecular clouds, CCMs were found to be abundant in dark and quiescent cores. TMC-1 is the typical carbon-chain-producing region. Molecules of \hcsn, HC$_9$N, and HC$_{11}$N
        were first detected in this core
        \citep{1978ApJ...219L.133K,1978ApJ...223L.105B, 1997ApJ...483L..61B}.
Almost all CCMs found so far, including C$_\mathrm{n}$O and C$_\mathrm{n}$S,
 were discovered in TMC-1 \citep {1984A&A...138L..13H, 1984Natur.310..125M,1988ApL&C..26..167I}.
 High-resolution maps of a number of S-bearing and N-bearing species were made and at least six cores were detected in the TMC-1 ridge \citep {1992ApJ...394..539H}. Changes in molecular abundances along the ridge were analyzed with the dynamical-chemical model \citep{2000ApJ...535..256M}.
 In addition to TMC-1, L1521B, L1498, L1544, and L1521E were also detected as rich starless carbon-chain-producing regions \citep{1992ApJ...392..551S,1996ApJ...468..761K,1999ApJ...518L..41O,2002ApJ...565..359H}. All these cores in the molecular complex Taurus are in an early stage of CCM chemistry.  
 Meanwhile a number of starless cores outside of Taurus, such as L492, Lupus-1A, L1512, and Serpens South 1a (Serp S1a), have been found to be abundant carbon-chain-producing regions \citep {2006ApJ...646..258H,2010ApJ...718L..49S,2011ApJ...730L..18C,2013MNRAS.436.1513F,2016ApJ...824..136L}.
Starless and dark cores CB 130-3 in the Aquila rift
region and L673-SMM4 in Cloud B of L673 
were also identified as carbon-chain-producing regions \citep{2011ApJ...736....4H}. Recently, 17 high-mass starless cores (HMSCs) and 35 high-mass protostellar objects (HMPOs) were surveyed with \hctn~ and \hcfn~ by \citet{2018ApJ...854..133T}. The molecule \hctn~ was detected in 15 HMSCs and 28 HMPOs, and \hcfn~was found in 5 HMSCs and 14 HMPOs \citep{2018ApJ...854..133T}.

 All these cores are in an early chemical phase. However, their properties and evolutionary states could be somewhat different.
In L1498, radial chemical differentiation has been detected with C$_2$S and NH$_3$ distributed in an onion shell-like structure with NH$_3$ at the inner part and C$_2$S at the outer part, which may have resulted from a slowly contracting dense core with a growing outer envelope \citep{1996ApJ...468..761K}. A similar structure of C$_2$S emission was discovered in L1544, which was interpreted as a result of infall and rotation by  \citet{1999ApJ...518L..41O}. Their different evolutional states can be sensitively traced with abundance ratios of
CCMs and NH$_3$ \citep{1988A&A...196..194O,1992ApJ...392..551S,2006ApJ...646..258H}.


In star-forming cores, CCMs are usually less abundant than in early cold and dark cores \citep{2008ApJ...672..371S}. In particular, the abundance of S-bearing species is significantly lower in protostellar cores \citep{1992ApJ...392..551S}.

However, in 2008 the protostellar core L1527 containing an infrared source IRAS 04368+2557 was found as a CCM-harboring region \citep{ 2008ApJ...672..371S}. { High-energy excitation} lines \citep[upper level energy $E_{up}$$>$20 K;][]{2008ApJ...672..371S} of
    carbon-chain molecules such as \cfht~($10_{0,10}-9_{0,9}$), \cfh~$N=9-8$,
    $l$-C$_3$H$_2$ ($4_{1,3}-3_{1,2}$), and \chtcch~$J=5-4, K=2$ were detected in this source. The intensity
    $(T_{MB})$ of the line \cfh~$N=9-8$, $J=19/2-17/2$ reaches 1.7 K. 
    These results are unusual since CCMs are generally absent in star-forming regions.
    A hypothesis was proposed to explain these findings. Since the evaporation temperature of CH$_4$ is about 30 K, it can be abundant in some warmer regions around protostellar
    objects \citep{2009ApJ...697..769S}. Subsequently, CH$_4$ reacts with C$^+$  to form hydrocarbon ions.
      Such processes were proposed as warm carbon-chain chemistry (WCCC) by \citet{2008ApJ...672..371S}, which is different from the chemistry in early cores.
       More recently, formation of CCMs in WCCC (lukewarm corinos) was modeled with the macroscopic Monte Carlo method and it was found
that the amount of CH$_4$ can diffuse inside the ice mantle, and therefore sublimation upon warm-up plays a crucial role in the synthesis of
carbon-chain species in the gas phase \citep{2019A&A...622A.185W}.

    A second WCCC source,  IRAS 15398-3359, was found by \citet{ 2009ApJ...697..769S} very shortly after the discovery of L1527. A massive star-forming region
    NGC 3576 was also revealed as a WCCC source according to the detected C$_5$H $J=39/2-37/2$ \citep{ 2015ApJ...798..36S}.

     In addition to high-energy excitation lines of hydrocarbons, other kinds of CCM emission lines exist in WCCC sources.
     N-bearing CCMs have been detected in all the WCCC sources found so far.
     Spectral lines of \hctn, \hcfn, \hcsn,~and even \hcnn~have been seen in L1527. Furthermore,
     \hctn~ J=10-9 and
     \hcfn~$J=32-31$, the latter being a very-high-energy excitation transition,
     were detected in the second WCCC source \citep{2008ApJ...672..371S,2009ApJ...697..769S,2015ApJ...798..36S,2012MNRAS.419..238B}.
     \hctn~$J=11-10$ was detected in NGC 3576 \citep{ 2015ApJ...798..36S}.

     L1489 is a famous low-mass star-forming source located in Taurus with a distance of 140 PC \citep{1988ApJ...324..907M}.
    In this paper we report CCM emissions detected towards the L1489 starless core, that is, the eastern molecular core (EMC) of the L1489 IRS  \citep{1989ApJS...71...89B,2002ApJ...572..238C}.
We made observations of multiple spectral lines of CCMs for this core, including emissions of N- and S-bearing  CCMs in the K$_u$ band and high-energy excitation transitions in the 3 mm band.

The observations are described in the following section. In Sect. 3 we present the results. The discussions are presented in Section 4 and a summary is given in Sect. 5.

\section{Observation} \label{sec:obs}
First we observed the spectral lines toward R.A.(J2000)= 04:04:47.5, Dec.(J2000)=26:19:42, which is the center of the high-visual-opacity region \citep{1989ApJS...71...89B} of L1489 (hereafter O point, which is also taken as the coordinate reference position in this work). We then mapped this source.

The lines we observed include the transitions of N-bearing and S-bearing species \hctn~$J=2-1$, \hcfn~$J=6-5$, \hcsn~$J=14-13$, $J=15-14$, $J=16-15$ as well as \cts~$J=3-2$ in K$_u$ band,
and  \cfh~$N=9-8$, $J=19/2-17/2$, $J=17/2-15/2$, \chtcch~$J=5-4$, K=0, $J=5-4$ K=1, and $J=5-4$, K=2, c-C$_3$H$_2$ ($2_{1,2}-1_{0,1}$) as well as \hctn~$J=10-9$ in the 3 mm band.
The parameters of the observed transitions listed in Table \ref{table_freq} are quoted from the molecular database at ``Splatalogue'' \footnote{www.splatalogue.net} that is a compilation of the Jet Propulsion Laboratory \citep{1998JQSRT..60..883P}, Cologne Database for Molecular Spectroscopy (CDMS; \citep{2005JMoSt.742..215M}), and Lovas/NIST \citep{2004JPCRD..33..177L} catalogs.
 This source was searched for \hctn~$J=4-3$, $J=5-4$, \hcfn~$J=8-7$, $J=17-16$ as well as C$_2$S
 $J_N=2_1-1_0$, $J_N=4_3-3_2$ \citep{1992ApJ...392..551S,1993ApJ...418..273F,2007A&A...461.1037B}.
All the transitions in this work were observed for the first time.

  \subsection{Observed with TMRT 65 m telescope}
The spectral lines at K$_u$ band were observed with the Tian Ma Radio Telescope (TMRT) of Shanghai Observatory  on Jan 25, 2016, and Dec 2-4,  2017.
The TMRT is a newly built 65 m diameter fully steerable
radio telescope located in the western suburb of Shanghai \citep{ 2016ApJ...824..136L}. The front end of the K$_u$ band is a cryogenically cooled receiver covering the frequency range of
        11.5$-$18.5 GHz. The pointing accuracy is better than 10\arcsec.
    An FPGA-based spectrometer
        based upon the design of Versatile GBT Astronomical Spectrometer
        (VEGAS) was employed as the Digital backend system (DIBAS)
        \citep{2012AAS...21944610B}. For molecular line observations,
DIBAS supports a variety of observing modes, including 19
single sub-band modes and 10  modes with eight sub-bands each.
        The center frequency of the sub-band is tunable
to an accuracy of 10 kHz. For our observation, the DIBAS mode
        22 was adopted.  Each of the eight side-bands
has a bandwidth of 23.4 MHz and 16384 channels.
The main beam efficiency is 60$\%$ at
        the K$_u$ band \citep{2015AcASn..56...63W,2016ApJ...824..136L}.
    The beam sizes and the
        equivalent velocity resolutions are given in the last
        two columns of Table \ref{table_freq}, respectively.
After the spectra at O point were observed, we made nine-point mapping observations on Dec 2-3, 2017.
The observations were performed in point-by-point mode around a point 1\arcmin~south of the O point to cover the L1489 IRS.
We start the map in a square pattern and with a grid separation of 1$\arcmin$; its diagonal lines are along E-W and N-S directions.
To cover the northern part of the emission region, we added four sampling points: (1,0), (0,0.5), (0,-0.5), and (-0.5,0) on Dec 4, 2017.

    \subsection{Observed with the Purple Mountain Observatory  telescope }

The spectral lines at 3 mm were observed with the 13.7 m telescope of the Qinghai Station of the Purple Mountain Observatory (PMO) on  March 31, 2017. 
The pointing and tracking
accuracies were both better than 5$\arcsec$. The main beam efficiency is 59$\%$  \footnote{\url{http://www.radioast.nsdc.cn/zhuangtaibaogao.php}}.
A Superconducting Spectroscopic Array Receiver with sideband separation was employed at the front end \citep{2012PhRvB..86l5303S}
while at the back end a Fast Fourier Transform Spectrometer with a total bandwidth of 1 GHz allocated to 16,384 channels was used.
The observed lines were 
covered in the two sidebands with frequencies from 85 to 86 GHz and 90 to 91 GHz, respectively. The spectral
resolution is about 61 kHz. 
 The system temperatures are in the range of
139$-$149 K, with a mean value of 144 K.
The position-switch mode was adopted. 
The on-source time was about 25 minutes for all observed lines except CH$_3$CCH J=5-4 which took an integrating time of about 20 hours and was completed on Dec 10-14, 2018. The maps were carried out with the on-the-fly mode on Nov 1, 2017.
The mapping region is 15$\arcmin$$\times$15$\arcmin$ for the lines of the CCMs in the 3 mm band with a 20$\arcsec$$\times$20$\arcsec$ grid. 

 The IRAM software package GILDAS including CLASS and GREG
        was used for all the line data reduction \citep{2000ASPC..217..299G}.

\section{Results} \label{sec:results}

All the transitions in K$_u$ band and 3 mm band are detected towards the EMC.
Hyperfine lines of \hctn~$J=2-1$ and \hcfn~$J=6-5$ are detected.
The velocity separation between $F=7-6$ and $F=6-5$ of \hcfn~$J=6-5$ is $\sim$0.3 km/s.
\citet{2004PASJ...56...69K} detected \hcfn~$J=6-5$ in TMC-1 with the Nobeyama 45 m radio telescope.
The hyperfine structures were not resolved at that time due to their slightly poorer spectral resolution ($\sim$0.7 km/s).
\citet{2016ApJ...824..136L} detected this line towards Serpens South 1a with TMRT 65 m. However,
only  the \hcfn~$J=6-5$, $F=5-4$ was resolved  because the
line width of Serpens South 1a is about 0.5 km/s.
Three hyperfine components $F=7-6$, $F=6-5,$ and $F=5-4$ of \hcfn~$J=6-5$ are fully resolved  for the first time.

 \subsection{Emissions of HC$_3$N $J=2-1$}

\begin{figure*}[htbp]
        \setlength{\abovecaptionskip}{10pt}
        \setlength{\belowcaptionskip}{-18pt}
        \centering
        \includegraphics[width=0.8\linewidth]{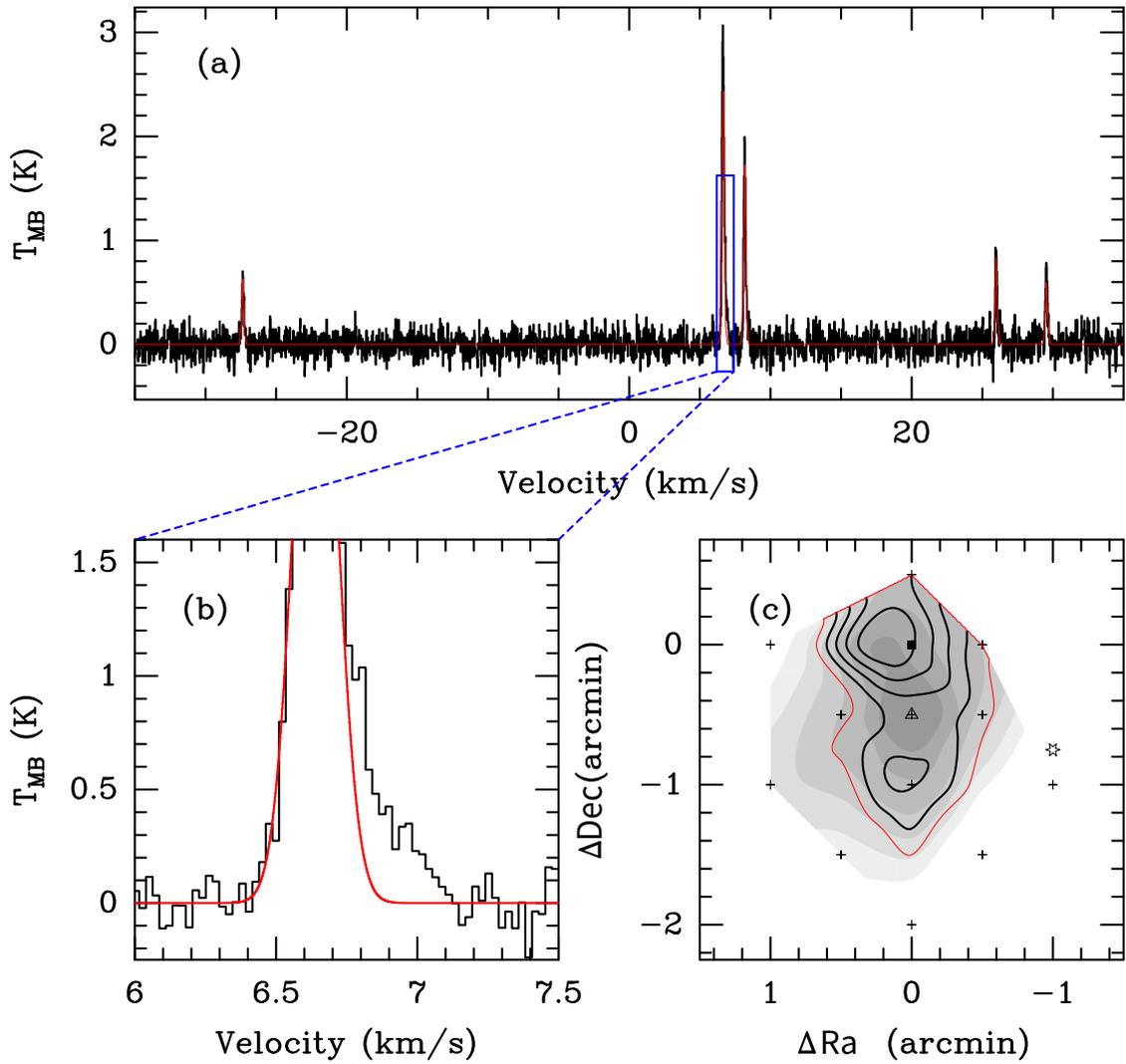}
     \caption{ (a): \hctn~J=2-1, F=3-2 spectrum of the peak position. Red lines show the Gaussian fitting. (b): Zoom-in of panel (a) showing line wing of HC$_3N$ J=2-1, F=3-2.
      (c): Background is integrated map of \hctn~J=2-1, F=3-2. Black contours represent integration of red wing (6.75 km/s $-$ 7.08 km/s) of \hctn~J=2-1, F=3-2 stepped from
      60 to 90 $\%$  by 10 $\%$  of maximum value 0.14 K km/s.  The red contour denotes 3 $\sigma$ level (0.06 K km/s) of \hctn~wing integration. The filled black
      square represents O point, and the black triangle represents P point (see text). 
      The IRS is shown by the hexagonal star \citep{2014ApJ...793..1Y}.  Small black crosses show sampled points of K$_u$ band observation.
      \label{figure_wing}
      }
 \end{figure*}

    Panel (a) of Figure \ref{figure_wing} presents the \hctn~$J=2-1$ spectrum and its wing at the peak position of the \hctn~map. In total, five hyperfine lines of \hctn~$J=2-1$ are well resolved, with the strongest one labeled as $J=2-1$, $F=3-2$
    with a T$_{MB}$ as high as 3.23 K.
   When zooming in on the hyperfine component \hctn~$J=2-1$, $F=3-2$ shown in panel (b),  it is clear to see that
        the residual spectrum after removing the Gaussian component
        displays a red wing.
The velocity of the shifted gas has a width of 0.33 \kms~spanning from 6.75 to 7.08\kms.
Panel (c) shows the contours of the integrated intensity of the red wing with a maximum value of 0.14 K \kms~and a $\sigma$ of 0.02 K \kms  \ overlaid on the map of the integration of line center. The red wing may belong to high-velocity gas since the profile of the wing is rather smooth and the ratio of the wing range to the FWHM of the line is similar to that of the red wing of the molecular outflow S140 \citep{1985ARA&A..23..267L}. However, foreground and background cold gas as well as additional components cannot be excluded \citep {2005AJ....129..330W}. High-resolution observations may be useful to identify its origin. Further discussions are excluded in the following analysis.

 \subsection{Spectral lines and emission regions} 
  The spectral lines of \hcfn~$J=6-5$ and \hcsn~$J=14-13$, $J=15-14$, $J=16-15$ as well as the \cts~$J=3-2$ at the map peak position (P point, see below) are shown in the left panel of Fig. \ref{figure_P_Ku}; the spectrums of \hctn~$J=2-1$, $F=3-2$   are presented in Fig. \ref{figure_wing}(a). The spectral lines of \cctht~(2$_{1,2}$-1$_{0,1}$), \chtcch~$J=5(0)-4(0),$ and \hctn~$J=10-9$ in the 3 mm band at the P position (see below) are presented in the left panel of Fig. \ref{figure_P_3mm}. The K=0, 1, and 2 of \chtcch~$J=5-4$ were all detected using PMO with a long integration time and are presented in the left panel of Fig. 5. The signal-to-noise ratio (S/N) of the weakest K=2 hyperfine line is larger than 5.

All the spectral lines were fitted with a Gaussian function.
The line parameters including the central velocity V$_\mathrm{LSR}$, the main beam temperature $T_\mathrm{MB}$,
the full width at half maximum (FWHM), and integrated intensities are listed in columns 3-5 of Table \ref{table_derived}.
One can see that all the lines have a V$_{LSR}$ of about 6.6 \kms~ except for \cfh~and \cctht\  whose V$_{LSR}$ are about (6.8-6.9) \kms. The FWHMs of these two lines (\cfh~and \cctht) are also broader than those of other detected CCM lines.

Integrated intensity maps of all the detected lines were made. The maps of the lines in K$_u$ and the 3 mm band are presented in the right panels of Figs. \ref{figure_P_Ku} and  \ref{figure_P_3mm}, respectively.

 One can see from the right-hand panels of Figs. \ref{figure_P_Ku} and \ref{figure_P_3mm} that the emission peaks (black hollow triangle) of transitions of N-bearing species in K$_u$ band are relatively consistent with each other. This peak point (P point hereafter) is located at R.A.(J2000)= 04:04:47.5, Dec.(J2000)=26:19:12,  about 0.5 arcmin south of the O point (black squire) and 1 arcmin east of the IRS (hexagonal star). The emission peaks of other lines are all with some deviations from the P point. For example, the peak of \cctht~(green filled hollow triangle in Fig. \ref{figure_P_3mm}) is located at $<$ 0.5 arcmin south of the P point. These deviations are due to different chemical and dynamical  properties of different molecular species and can often be seen in molecular cores such as the NH$_3$ cores in Orion, Cepheus, and H$_2$O maser sources as well as various gas cores in the carbon-chain-producing region NGC 3576 \citep[see also Fig. \ref{figure_continuum} and Sect. 4.1]{1993A&AS...98...51H,2006A&A...450..607W,2015ApJ...798..36S}. We take the P point as the peak of the emissions of all the CCMs  except \cts~$J=3-2$.

\subsection{Lines at O point}
Single point observations towards O point were made before the map. The highest S/Ns of the data at this point were compared to those of other points. The comparison enabled us to analyze the profile of molecular lines in  detail.
The lines at O point are thus also shown in Fig. \ref{figure_O}. The left panel of Fig. 4 presents the spectral lines of \hctn, \hcfn, \hcsn,~and \cts~in the K$_u$ band. The right panel  presents the spectral lines of \cfh, \chtcch, \cctht~, and \hctn~in 3 mm band.

The spectral peak of the \hcsn~$J=14-13$ at O point shown in Fig. \ref{figure_O} seems to have a dip at the center with an S/N $\approx$3. The spectrum of \hcsn~$J=15-14$ may also be split but interference of noise cannot be excluded. This was not seen in the spectrums of other CCMs in the K$_u$ or 3 mm bands at O point. For the emission of \cts~$J=3-2$, the T$_{MB}$ at O point is higher than that at P point, and therefore O point is adopted as the peak of the \cts~core.

\subsection{Column density}
The column densities of the observed CCMs except \cts~were calculated from the integrated intensities of lines of P point. For \cts, parameters at O point are adopted. Assuming the gas is in local thermodynamic equilibrium (LTE) and the lines are optically thin, the column densities are calculated with the solution of the radiation transfer equation \citep{1991ApJ...374..540G,2015PASP..127..266M}
\begin{equation}
 N=\frac{3k}{8\pi^3\nu}\frac{Q}{\sum S_{ij}\mu^2}
exp\left(\frac{E_{up}}{kT_{ex}}\right) J(T_{ex})
  \int \tau d V \label{Eq_N}
,\end{equation}
                \begin{eqnarray} \label{eq_tex}
                        T_{MB}/\eta =f\frac{h\nu}{k} [({e^{\frac{h\nu}{kT_{ex}}}-1})^{-1}-({e^{\frac{h\nu}{kT_{bg}}}-1})^{-1}][1-e^{-\tau}]
        ,\end{eqnarray}
        where S$_{ij}$, $\mu$, and $Q$ are the line strength,
        the permanent dipole moment, and the partial function, respectively, which are quoted from ``Splatalogue'', and T$_{MB}$ is the main beam temperature, $\eta$ is the efficiency of the main beam, and $f$ is the beam filling factor, assumed to be unity.

The excitation temperature was calculated using different methods in this work for comparison.
Using the hyperfine structure (HFS) fitting program in GILDAS/CLASS, we performed hyperfine structure fitting toward spectra of \hctn~$J=2-1$.
The optical depth of the main component was obtained as $\tau=0.45\pm 0.12$ and an excitation temperature of $11.5\pm 1.0$ K was derived
through Eq. \ref{eq_tex} with T$_{\rm bg}$=2.73 K and a beam filling factor of  unity.
We use the lines K=0, 1, and 2 of \chtcch~ $J=5-4$ to derive the rotational temperature T$_{rot}$ and a value of 12.6$\pm 1.0$ K was obtained (see the middle panel of Fig. 5). Since the K=1,2 levels of \chtcch~ J=5-4 are easily thermalized and even more easily than NH$_3$ (1,1) (2,2), the kinetic temperature T$_k$ is assumed equal to T$_{rot}$, 12.6$\pm 1.0$ K
\citep{1984A&A...130..311A,1980IAUS...87....1E}.

Continuum data from far-infrared to sub-millimeter are available for the L1489 region \footnote{\url{http://www.cosmos.esa.int/web/herschel/science-archive}}. The dust temperature (T$_d$) and H$_2$ column density were derived via modeling the PACS and SPIRE data of Herschel at  160, 250, 350, and 500 $\mu$m (see Table 3 and the right panel of Fig. 5) to a modified black body:
\begin{equation}
S_\nu=B_\nu(T)(1-e^{-\tau_\nu})\Omega,
\end{equation}
where $\tau_\nu=\mu_{H_2}m_H\kappa_\nu N_{H_2}/R_{gd}$ ; here $\mu_{H_2}$ = 2.8 is the mean molecular weight adopted from \citet{2008A&A...487..993K}, $m_H$ is the mass of a hydrogen atom, $N_{H_2}$ is the
column density, and $R_{gd}$ = 100 is the ratio of gas to dust.
The dust opacity $\kappa_\nu$ can be expressed as a power law of frequency,
\begin{equation} \kappa_\nu=5.9({\nu}/{850\ GHz})^\beta$ cm$^2$ g$^{-1}, \end{equation}
with  $\kappa_\nu$(850 GHz) = 5.9 cm$^2$ g$^{-1}$ adopted from \citet{1994A&A...291..943O}. The free parameters are the dust temperature, dust emissivity index
$\beta,$ and column density.
The fitting results give T$_d$=13.8$\pm$0.2 K and $N_{H_2}$=(1.02$\pm$0.07)$\times$10$^{22}$ cm$^{-2}$ with $\beta=1.75$.

These derived temperatures are listed in Table 4. The T$_{d}$ variation from the inner to outer parts of the core, which was model fitted from the SCUBA 850 $\mu$m image \citep{2011ApJ...728..144F}, is also listed in Table 4. These values are comparable with the kinetic temperature of L1527 (12.3$\pm$0.8 K) and
that of IRAS 15398-3359 \citep[12.6 K;][]{2008ApJ...672..371S,2009ApJ...697..769S}.
 Therefore, a unified excitation temperature of 12.6 K was adopted for calculating column densities and for further analyses.

For the column density of the detected species, only that of \chtcch~can be derived by analysis of multiple transitions at different energy levels. For other species including the hyperfine lines of \hcfn~ and multiple rotation lines of \hcsn~, the upper level energies of the multiple lines are close to one another and this method is impractical. The average values of column densities calculated from multiple lines weighted by T$_{MB}$/$\sigma$ was therefore adopted as the species column density.

The T$_{ex}$ uncertainty will bring in about 10$\%$ error for calculation of column densities of all the species.

The column densities are listed in Table \ref{table_derived}. The uncertainties 
of column densities are derived from errors of T$_{ex}$ and line integrated intensities.
The column densities of our N-bearing molecules range from $1.4\times 10^{12}$ to $4.5\times 10^{13}$ cm$^{-2}$.
Among all the detected molecules in the K$_u$ band,  \cts~has the lowest column density, 0.8 $\times$ 10$^{12}$ cm$^{-2}$, which is much lower than those of starless cores such as TMC-1, L1544, and L1498 \citep{ 1992ApJ...392..551S}; this is also lower than that of the stellar core L1251A \citep{2011ApJ...730L..18C}.

\section{Discussion} \label{sec:obs}

\subsection{Carbon-chain molecule emission characteristics}

 The species \cfh~N=9-8, J= 19/2-17/2, 17/2-15/2 and \chtcch~ J=5-4, K=0, 1, 2 were detected in the WCCC source L1527 \citep{2008ApJ...672..371S}. \chtcch~ J=5-4, K=2 is the highest-excitation line detected in L1527. All of these lines were detected in the EMC. In the EMC, N-bearing species appear to be relatively abundant while \cts~ is very weak.

Figure 6 presents
the CCM column densities of the EMC and L1527 as well as the five early carbon-chain-producing regions normalized by the values of TMC-1.
 One can  see that the column densities of {the species with high-energy excitation lines} in the EMC and L1527 are close to each other. The \chtcch~ column densities of the two cores are slightly higher than that of TMC-1 while their values of \cfh~are lower than that of TMC-1. Among five starless cores, L1521B is the only one{ where the \cfh~is detected}. It is worth noting that the column density of \cfh~in L1521B is even higher than that of TMC-1. In this source the detected \cfh~ line is $N=2-1$ $J=5/2-3/2$, $F=2-1, 3-2$ with an upper level  energy of 1.4 K, which is much lower than those of lines detected in the EMC and L1527 \citep{2004ApJ...617..399H}. However, the \hctn~ J=10-9 with E$_{up}$ 24.0 K was detected in the starless core L1498 \citep{2006A&A...455..577T}. Since only the EMC and L1498 have this transition observed, it was not plotted on Fig. 6.

From Fig. 6 one can also see that among all the starless cores, the column densities of the N-bearing species in the EMC and L1527 are at the intermediate level. The \hctn~ column density in the EMC is higher than that of L1527. The N-bearing species seem to be abundant in WCCC sources in general. In the WCCC source IRAS 15398-3359, a very-high-energy excitation line of \hcfn~ J=32-31 with E$_{up}$ 67.5 K was detected though the detection is tentative \citep{2009ApJ...697..769S}. The \hctn~ column density in IRAS 15398-3359 is 1.5x10$^{14}$ cm$^{-2}$ (Wu et al. in preparation).  In the massive core WCCC source NGC 7536 the \hcfn~ column density is 1x10$^{13}$ cm$^{-2}$ which is derived from the detected line of \hcfn~ J=11-10 \citep{2015ApJ...798..36S}. The E$_{up}$ of the \hcfn~ J=11-10 transition is 28.8 K.

As for the S-bearing species, Fig. 6 shows that the column density of the EMC is a little lower than that of L1527.  However, \cts~column densities of both sources, about 6.5\% (EMC) and 10\% (L1527) of the TMC-1 value, respectively, are much lower than all of the compared early starless cores.
It was recognized previously that S-bearing species are usually deficient in star-forming cores.
A number of low-mass star-forming cores were examined for emissions of S-bearing CCMs by \citet{1992ApJ...392..551S}. Results showed no or marginal detection for most protostellar cores  and L1489 was taken as an example. 
In the EMC, the emission of \cts~$J=2-1$ at P point is also marginally detected.
Our C$_3$S $J=3-2$ map shows that the emission peak of \cts~$J=3-2$ (O point) is separated from those of N-bearing species and NH$_3$ \citep {1988ApJ...324..907M}, similar to the cases of L1498 and L1544. But even at the O point, the column density of the \cts~ is still much lower than those of L1544 and L1498.

All these comparisons present the following characteristics of the EMC.

1. Emissions of {the species with high-energy excitation lines} of the EMC are close to those of L1527. For these two sources
the column densities of {the species with high-energy excitation lines} are comparable with those of TMC-1.

2. For the N-bearing species, the column densities of the EMC and L1527 are at the medium level among all the samples.

3. The column density of \cts~ of the EMC is close to that of L1527 but is much lower than those of all starless cores.

In short, the CCM emissions of the EMC are similar to those of L1527 but deviate from the early cold starless cores.

 \subsection{Core status and environment} \label{diss_3}
L1489 is one of the 90 small visually opaque regions chosen from the Palomar Sky Atlas prints \citep {1983ApJ...264..517M}. 
NH$_3$ (1,1) mapping presents a gas core which is about half an arcminute south of the high visual opacity and about one arcminute east of the IRAS 04016+2610, which was revealed as a Class I IRS \citep{1988ApJ...324..907M,1987ApJ...319..340M}. This region was also mapped with N$_2$H$^+$ J=1-0 \citep{2002ApJ...572..238C}. The peak positions of the NH$_3$ (1,1) and N$_2$H$^+$ J=1-0 map are closely coincident with our P point (1 arcmin east of L1489 IRS).
 These observations confirm that the EMC is dark and starless, similar to those early starless cores presented in Fig. 6.

The C$_2$S distributions of L1498 and L1544 have a central hole which can be explained with infall and rotation \citep{2006ApJ...646..258H,1999ApJ...518L..41O}.  In the EMC no sign of collapse or rotation has been detected so far and it is quiescent. At P point there is strong NH$_3$ emission together with weak
\cts~emission, which may be related to the evolutional state
of the core \citep{1992ApJ...392..551S,2006ApJ...646..258H}.

The abundance ratios of NH$_3$/C$_2$S of TMC-1, L1521B, L492, L1498 and L1544 range from 2.9 to 25 \citep{2006ApJ...646..258H}. 
 For the EMC, using the NH$_3$ column density of \citet{1988ApJ...324..907M} and the C$_2$S column density estimated from \cts~emission with an average abundance ratio
 C$_2$S/\cts~of 4.3$\pm$0.9 derived from the data of starless cores including L1498,
 L1521B, 
TMC-1 
and L1544 
\citep{1992ApJ...392..551S},
the ratio of  NH$_3$/C$_2$S  is 289.
This is one order of magnitude larger than the largest of the seven cores listed in \citet{2006ApJ...646..258H}. It is also larger than the NH$_3$/C$_2$S ratio of 37 for Serp S1a, which was derived from  NH$_3$ and \cts~ column densities with the ratio of C$_2$S/C$_3$S 4.3 \citep {2013MNRAS.436.1513F,2016ApJ...824..136L,1992ApJ...392..551S}.
Serp S1a has a more active and complex environment than starless cores in Taurus and infall was detected \citep{2013MNRAS.436.1513F}.
These indicate that the EMC has the latest evolutionary state  among all the starless cores shown in Fig. 6.

A prominent difference between the EMC and other early starless cores is the temperature and the thermalisation of the gas.
In TMC-1 the rotational temperature ranges from 4 to 8 K while the kinetic temperature ranges from 9 to 10 K 
 \citep{1991PASJ...43..607K,2004ApJ...610..329K, 2006ApJ...647..412S,2008ApJ...672..371S}.
The dust temperature is 10.5-12 K \citep {2016A&A...590A..75F} and 10.6$\pm$0.1 K (Wu et al. in preparation). 
Serp S1a has a T$_k$ 10.8 K derived from NH$_3$ (1,1) (2,2) and 7 K was adopted as T$_{ex}$ \citep{2013MNRAS.436.1513F,2016ApJ...824..136L}. The
T$_{ex}$  of L492 was derived as 6.4 K from \hctn~ and 8.3 K from CO, respectively \citep{2006ApJ...646..258H}.
L1521B has a T$_{ex}$ of 5.5-6.5 K and its T$_K$ is about 10 K \citep{2004ApJ...617..399H,1992ApJ...392..551S}.
The T$_{ex}$ of \cts~ of L1498 ranges from 5.5 to 10 K \citep{1992ApJ...392..551S, 1996ApJ...468..761K}, and the T$_k$ is 10 K derived from NH$_3$ emission in this core. The T$_d$ of L1498 is 10 K \citep{2004A&A...416..191T}. The T$_{ex}$ of L1544 is from 5 to 6.5 K \citep{1992ApJ...392..551S} and T$_K$ derived from NH$_3$ is $\sim$ 10 K , and the T$_d$ of L1544 is also 10 K  \citep{2002ApJ...569..815T}. In the EMC however, the T$_{ex}$ is 11.5 K derived from \hctn~ and T$_{rot}$ is 12.6 K from \chtcch, which are similar to the T$_{rot}$  of  C$_4$H$_2$  (12.3$\pm$0.8 K)  and \chtcch~(13.9 K) for L1527 \citep{2008ApJ...672..371S}. These comparisons show that
the gas of the EMC is warmer and closer to thermalization status than those in early starless cores and comparable to that of L1527. These may explain why the
carbon-chain molecular emissions in the EMC are similar to those of WCCC sources.

However, there is a fundamental difference in that L1527 contains a protostar which is the heating source of the core material \citep{1974ApJ...189..441G,2008ApJ...672..371S}. 
For the EMC, there is no protostar inside, only an association with a protostellar object L1489 IRS at one arcminute to the west.

Figure 7 displays the images of 850 $\mu$m, 500 $\mu$m, and 250 $\mu$m continuum emissions as well as  molecular line emissions in the L1489 region.
 One can see that the P point (black triangle), the centers of larger 850 $\mu$m continuum core, and the NH$_3$ core (blue triangle) are close to each other but not overlaid exactly. Such deviation among peaks of different molecular line emission regions are very common in prestellar and stellar cores even in the TMC-1 region \citep {2014A&A...569A..11S,2016ApJ...833...97K, 1997ApJ...486..862P}.
Cores of other molecular lines such as
N$_2$H$^+$ J=1-0 and CS J=3-2 are also coincident with the EMC \citep{2002ApJ...572..238C,1989ApJ...346..168Z}.
The weaker but larger core in the east of L1489 IRS shown
in the SCUBA 850 $\mu$m map and the SPIRE 500 $\mu$m image covered
the EMC entirely.

The L1489 IRS is a Class I source while the one in L1527 is a Class 0 object.
The L$_{bol}$ is 3.6-3.8 L$_{\odot}$  for the L1489 IRS and 1.3-1.9 L$_{\odot}$ for L1527, respectively (Kristensen et al. 2012; Chen et al. 1995).
The T$_{bol}$ of L1527 is 44 K (Kristensen et al. 2012; Chen et al. 1995), while the T$_{bol}$ of the L1489 IRS is 200-283 K and more than five times of that of L1527 on average \citep{2012A&A...542A...8K,1995ApJ...445..377C}. The \cfh~ emission extends from the center
part of the envelope to 5600 AU in L1527.
High-energy excitation CCM lines in the EMC are possible to be excited, though L1489 IRS is located at the margins of the EMC since L1489 IRS is has relatively high bolometric luminosity and temperature. Further more, H$_2$O 1$_{1,0}$-1$_{0,1}$
observed with HIFI on Herschel presents an inverse P-cygni profile in L1527 but a P-Cygni profile in L1489 IRS, indicating that  L1489 IRS lacks an outer, cooler gas layer, which is in favor of the heating of its neighboring material.

From Fig. 7 one can see that at each wavelength, the strongest emission is around the IRS. The higher the frequency, the more the contours concentrate at the IRS. The tails of the contours of the 250 $\mu$m and 500 $\mu$m maps present a trajectory format starting from the IRS, which has not previously been seen in low-mass protostellar cores.
These may be the evidence that the EMC is heated by the L1489 IRS. The size of the 250 $\mu$m emission region is 169.1 arcsec and T$_d$  from the spectral energy distribution (SED) fitting based on data of  PACS-SPIRE bands is 13.8 K in the whole region.


The heating of L1489 IRS can be more significant considering the possible evolutionary history of the relationship between L1489 IRS and the EMC; they may be neighbors since the birth of L1489 IRS.
However, another possibility may be that the IRS has moved away from the EMC center to the cloud margin,
or it has dispersed the surrounding gas and has separated itself from the EMC \citep{1993A&AS...98...51H,2006A&A...450..607W}.
The influences of the L1489 IRS might what gives the EMC its WCCC characteristics.

However, WCCC began from reactions of C$^+$ and sublimated CH$_4$ from dust. A precondition is that the temperature needs to be $\sim$30 K \citep{2009ApJ...697..769S}. For L1527, the kinetic temperature is 20-30 K derived from the \cctht~ 4$_{3,2}$-4$_{2,1}$ line detected with the Plateau
de Bure Interferometer \citep{2010ApJ...722.1633S}. From the \cctht~ and SO lines measured with ALMA with beam sizes 0.8\arcsec$\times$0.7\arcsec~and 0.7\arcsec$\times$0.5\arcsec~and analyzed with non-LTE large-velocity-gradient code, the kinetic temperatures of the \cctht~ emitting region was found as 30 K at 100 AU \citep{2014Natur.507...78S}. The kinetic temperatures of L1527 measured from the interferometers are much higher than 13.9 K derived from the \chtcch~ J=5-4, K=1,2 observed using  NRO 45 m telescope with beam size $\sim$20\arcsec. These results present effects of telescope beams. The kinetic temperature of the EMC derived from \chtcch~ J=5-4, K=0,1,2 detected with the PMO 13.7 m telescope with a beam size of 53\arcsec~is 12.6$\pm$1.0 K. The dust temperature fitted from JCMT and Herschel data is 13.8$\pm$0.2 K. Both the values of the kinetic and dust temperature are comparable to those of L1527 measured with the NRO 45 m telescope. For the IRS sources, both the L$_{bol}$ and T$_{bol}$ of the L1489 IRS are higher than those of L1527. These comparisons indicate that in a smaller region within the EMC the kinetic temperature may be higher than the current results. Higher-resolution observations are needed to measure the temperature.

In the WCCC source L1527, CCMs of the prestellar phase could survive if the prestellar collapse is faster than that of other star-forming cores \citep{2008ApJ...672..371S}.
The EMC should have an early and a cold phase, like those starless cores shown in Fig. 6. No collapse or infall have been found in the EMC so far. The CCMs detected in the EMC may have survived from its early phase.
In particular, the chemical activities of the CCMs seem to be related to their existence in the WCCC phase.
Because of the different polarities, the  N- and S-bearing species are more active than C$_4$H and CH$_3$CCH and more likely to react with their partners.
For example, the major loss route of C$_4$H is through 
reaction with C$^+$ \citep{1984MNRAS.207..405M}:
C$_4$H+C$^+$ $\rightarrow$ C$_5^+$+H, which has a rate coefficient of 2.0 $\times$10$^{-9}$ cm$^3$ at 10 K. While the reaction of HC$_3$N with C$^+$: HC$_3$N+C$^+$ $\rightarrow$ C$_3$HN$^+$+C has a rate coefficient of 8.7 $\times$10$^{-9}$ cm$^3$ at the same temperature \citep{1984ApJS...56..231L}.  However, both species are formed from hydrocarbon ions and the reaction rates are not significantly different from one  another \citep{1984ApJS...56..231L}. This suggests that {the molecular species with high-energy excitation lines} have a higher rate of survival than low-excitation lines, which might be a reason for the emissions of {species with high-energy excitation lines} in L1527 \citep{2008ApJ...672..371S} and the late core EMC. This may also be the reason for the detection of the \hctn~ J=10-9 line with E$_{up}$ 24.0 K in L1498 \citep{2006A&A...455..577T}.

Besides the high-excitation lines and the N-bearing and S-bearing species discussed above,  circular hydrocarbon \cctht~ with $T_{MB}$ 1.72 K was also detected in the EMC. The species \hctn~ J=10-9 with a high upper-level energy of 24.0 K is detected with $T_{MB}$ 0.63 K. Maps of all the detected species except for \cts~ show consistent emission peaks at about 1 arcmin east of the L1489 IRS. These results show that the EMC is an abundant CCM laboratory. Its comparable and contrasting conditions with the WCCC source L1527 and with early cold carbon-chain-producing regions may promote further CCM searches and help to constrain model analyses of CCMs.

\section{Summary}
Abundant carbon-chain molecules were detected toward the EMC of L1489 IRS, which is identified as a particular carbon-chain-producing region.

1. With the TMRT, \hctn~ J=2-1, \hcfn~ J=6-5, and \hcsn~ J=14-13, 15-14, 16-15 as well as  \cts~ J=3-2 were detected. The T$_{MB}$ of the hyperfine line J=2-1 F=3-2 of \hctn~ is 3.23 K. Five hyperfine lines of \hctn~$J=2-1$ were resolved. Hyperfine components of F=7-6, F=6-5, and F=5-4 of \hcfn~$J=6-5$ were resolved for the first time. The T$_{MB}$ of the three rotational lines of \hcsn~ ranges from 0.25 to 0.32 K.  
 The emission of \cts~J=3-2 is the weakest among the detected transitions. Using the PMO telescope,
 {high-energy excitation lines including C$_4$H N=9-8, J=17/2-15/2, 19/2-17/2,  CH$_3$CCH J=5-4, K=2, and \hctn~ J=10-9} were detected. The highest upper-level energy is 41.1 K.
All the transitions in K$_u$ and 3 mm band  are detected for the first time in the EMC of L1489 IRS.

2. Maps of the observed transition lines were also obtained. Emission peaks of our detected lines except \cts~are all located at about 1\arcmin~east of the L1489 IRS, which is consistent with previously detected gas cores and the 850 $\mu$m continuum east core. 

3. The CCM column densities of the EMC were compared with those of TMC-1 together with five carbon-chain-producing regions in early phase (Serp S1a, L492, L1521B, L1498 and L1544) and WCCC source L1527. Results show that the column densities of {the species with high-energy excitation lines including} \chtcch~and \cfh~in the EMC are close to those of L1527. The \cts~ column density of the EMC is slightly lower than that of L1527 but much lower than the five starless cores. The column densities of the N-bearing species are close to those of L1527, and the values of the both sources are at the intermediate level of the starless cores.

4. Similarly to early carbon-chain-producing regions, the EMC is dark, starless, and quiescent. However, the EMC is rather at a late evolutionary stage ($N$(NH$_3$)/$N$(\cts)=289), and is warmer than starless carbon-chain-producing regions. On the other hand, the temperature and thermalization of the EMC are close to those of L1527, though L1527 is a protostellar core. These indicate that the EMC is very special among the carbon-chain-producing regions detected so far.

5. The L1489 IRS has relatively high bolometric luminosity and temperature. The weaker but larger core to the east of L1489 IRS shown in the SCUBA 850 $\mu$m  and the SPIRE 500 $\mu$m images covered the EMC completely.
 The tails of the contours of the 250 $\mu$m and 500 $\mu$m maps present
a trajectory format starting from the IRS. The dust continuum emission area and the morphology of the contours show that the EMC is externally heated by the L1489 IRS.

\begin{acknowledgements}

        We are grateful to the staff of PMO Qinghai Station and SHAO.
    We also thank Shanghuo
        Li, Kai Yang and Bingru Wang for their assistance during the
        observation period. This project was supported by  the grants of National Key R\&D Program of China No. 2017YFA0402600, NSFC Nos. 11433008, 11373009, 11373026, 11503035, 11573036, U1331116 and U1631237, and the Top Talents Program of Yunnan Province.
This research used the facilities of the Canadian Astronomy Data Centre operated by the the National Research Council of Canada with the support of the Canadian Space Agency.
\end{acknowledgements}

\bibliographystyle{aasjournal}

\bibliography{ms}

\clearpage
\begin{table*}[h]
\setlength{\tabcolsep}{0.05in}
\caption{Observed transitions and telescope parameters \label{table_freq}}
\begin{tabular}{cccccccccc}
\hline\hline
Molecular  &Q$_{9.375}$ & Q$_{18.75}$ & Transition & freq.(MHz)  & $S_{ij}\mu^2(D^2)$ & $E_{low}(K)$ & $E_{up}(K)$ & FWHM(") & $v_{chan}(m/s)$\\
\hline
\hctn    &      43.27&86.22 &   J=2-1 F=1-1     &       18198.3745 &    2.311   &       0.43652 &        1.30990 &       52      &       24\\ 
\nodata  & & &  J=2-1 F=3-2     &       18196.3104              &       12.94   &       0.43667 &        1.30995 &       52      &       24\\
\nodata  & & &  J=2-1 F=2-1     &       18196.2169              &       6.933   &       0.43652 &        1.30980 &       52      &       24\\
\nodata  & & &  J=2-1 F=1-0     &       18195.1364              &       3.081   &       0.43681 &        1.31003 &       52      &       24\\
\nodata  & & &  J=2-1 F=2-2     &       18194.9195              &       2.312   &       0.43667 &        1.30988 &       52      &       24\\
\hcfn    & 147.1 & 293.8 &      J=6-5 F=7-6     &       15975.9831         &     43.3    &       1.91687 &        2.68359        &       60      &       27\\
\nodata  & & &  J=6-5 F=6-5     &       15975.9663              &       36.4    &       1.91673 &        2.68345 &       60      &       27\\
\nodata  & & &  J=6-5 F=5-4     &       15975.9336              & 30.6  &       1.91687 &        2.68359 &       60      &       27\\
\hcsn    & 346.7 & 693.0 &      J=16-15 &       18047.9697        & 325.3       &       6.49618 &        7.36235 &       53      &       24 \\
\nodata  & & &  J=15-14 &       16919.9791                      &       348.5   &       5.68414 &        6.49617 &       56      &       25 \\
\nodata  & & &  J=14-13 &       15791.9870                      &       371.7   &       4.92634 &        5.68424 &       60      &       27 \\
\cts     & 68.0 & 135.5 &       J=3-2   &       17342.2564  &   41.15   &       0.83234 &        1.66464 &       55      &       25 \\
\cfh     &      165.5 & 329.7 & N=9-8 J=19/2-17/2&      85634.00                &       13.36   &       16.436  &        20.546      &   53      &       210 \\
\nodata  & & &  N=9-8 J=17/2-15/2  &85672.57            &       11.94   &       16.451  &        20.563      &   53      &       210\\
\chtcch  & 34.42 & 88.27 & J=5(0)-4(0)  &               85457.27                &       2.81    &       8.20271 &        12.30399        &       53      &       210 \\
\nodata  & & & J=5(1)-4(1)  &           85455.62                &       2.70    &       15.40346&          19.50466   &   53      &       210  \\
\nodata  & & & J=5(2)-4(2)  &           85450.76                &       2.36    &       37.00571&          41.10667   &   53      &       210  \\
\cctht   & 72.4 & 201.8 & J=2(1,2)-1(0,1)&     85338.89      &   52.94    &   2.3498  &    6.44539    &   53  &   210 \\
\hctn    & 43.27 & 86.22  &  J=10-9      &   90979.0230          &   138.6   &   19.64854&    24.01482   &   53  &   210\\
\hline
\end{tabular}
\end{table*}

\clearpage
\begin{table*}
\setlength{\tabcolsep}{0.05in}
\caption{Observed and derived parameters \label{table_derived}}
\begin{tabular}{ccccccc}
\hline\hline
\colhead{Species}         & \colhead{Transitions}   &  \colhead{$V_{LSR}$}\tablefootmark{1}  & \colhead{$T_{MB}$}\tablefootmark{1}  & \colhead{$\Delta V$}\tablefootmark{1}  &  \colhead{$\int T_{MB}dv$}\tablefootmark{2}    &\colhead{$N(S)$ \tablefootmark{3}}    \\
                &               &  \colhead{(km/s)}     &  \colhead{(mK)}      &\colhead{(m/s)}      & \colhead{(K m/s)}           & \colhead{($10^{12}\ cm^{-2}$)}\\
\hline
\hctn           &       J=2-1 F=1-1     &       6.660(9)        &       692(110)        &       177(39) &       130(22)   &   \\
                &       J=2-1 F=3-2     &       6.650(2)        &       3230(110)       &       206(6)  &       709(18)  &   \\
                &       J=2-1 F=2-1     &       6.642(4)        &       2134(111)       &       164(11) &       372(7)    &   \\
                &       J=2-1 F=1-0     &       6.640(8)        &       1172(110)       &       145(18) &       172(6)    &    \\
                &       J=2-1 F=2-2     &       6.640(9)        &       852(110)        &       185(30) &       167(6)    & 45(4)\\
\hcfn           &       J=6-5 F=7-6     &       6.626(5)        &       628(46)         &       161(11) &       106(5)    &     \\
                &       J=6-5 F=6-5     &       6.625(5)        &       629(46)         &       194(13) &       126(5)     &    \\
                &       J=6-5 F=5-4     &       6.626(8)        &       427(46)         &       193(19) &       87(5)     & 9.7(6)\\
\hcsn           &       J=14-13 &       6.639(9)        &   257(30)              &       184(27) &  51(10)   & \\
                &   J=15-14     &       6.644(9)        &       318(30)         &       156(33) & 53(10)   &     \\
                &       J=16-15 &       6.601(7)    &   252(41)         &       168(18)  & 45(4)  & 1.4(0.2)\\
\cts            &   J=3-2   &  6.640(9)&   161(33)  &   151(19)& 26 (3)      &   0.8(0.1)\tablefootmark{4}   \\
\cfh            & N=9-8 J=19/2-17/2 &  6.79(7)  &  187(35)  &  520(120) &  122(24)     &      \\
                & N=9-8 J=17/2-15/2 &  6.78(3)  &  266(35)  &  535(87)  &  105(15)    &45(8)   \\
\chtcch         & J=5(0)-4(0)   &  6.74(4)  &  220(9)  & 359(155)    & 91(4)        &         \\
                & J=5(1)-4(1)   &  6.72(3)    &  170(9)  &  350(40)    &  83(8)         &         \\
                & J=5(2)-4(2)   &  6.72(3)    &  42(8)   &  310(40)    &  12(8)         & 20(3)        \\
\cctht         & 2(1,2)-1(0,1) &  6.77(4)    &  1720(40) &  563(15)    &  671(16)         & 5.5(0.1)    \\
\hctn & J=10-9 & 6.70(2)& 629(45) & 416(45) & 254(15)  & 3.4(0.3)\\
\hline
\end{tabular}
\\
\tablefoottext{1}{The uncertainties of $V_{LSR}$, $T_{MB}$ and $\Delta V$ are obtained from Gaussian fitting.}\\
\tablefoottext{2}{The uncertainties of $\int T_{MB}dv$ are calculated from error transfer formula.}\\
\tablefoottext{3}{Column densities of species are adopted as average values calculated from different hyperfine lines weighted by T$_{MB}$/$\sigma$, assuming all are optical thin with an identical excitation temperature, except that of \hctn~based on \hctn~$J=2-1$ whose optical depths are derived from HFS fitting and that of CH$_3$CCH whose column density is adopted as the value given by rotation diagram. Uncertainties of column densities are calculated from error transfer formula, including the error introduced from the uncertainty of excitation temperature.

 }\\
\tablefoottext{4}{\cts~parameters at O point are listed.}
\end{table*}

\begin{table*}
\setlength{\tabcolsep}{0.1in}
\caption{Dust parameters}
\begin{tabular}{ccccccccc}
\hline\hline
        \colhead{Source} & \colhead{S$_{70}$}& \colhead{S$_{160}$}& \colhead{S$_{250}$}&
\colhead{S$_{350}$}& \colhead{S$_{500}$}& \colhead{Size$_{250}$} &\colhead{T$_\mathrm{d}$} &\colhead{N$_\mathrm{H_2}$}\\
& \colhead{Jy} & \colhead{Jy} & \colhead{Jy} & \colhead{Jy} & \colhead{Jy} & \colhead{arcsec}
& \colhead{K} & \colhead{$10^{22}$ cm$^{-2}$}\\
\hline
      L1489        &    38.07     &    84.69   &    102.86    &    75.58     &   38.22 &   169.07 &    13.8(0.2)  & 1.0(0.1)\\
\hline\end{tabular}
\\
\tablefoottext{The 850 $\mu$m flux (5.78 Jy) is combined for the SED of L1489, which is from \citep{2000ApJ...534..880H}.}\\
\end{table*}

\begin{table*}
\setlength{\tabcolsep}{0.1in}
\caption{Temperatures}
\begin{tabular}{cccc}
\hline\hline
\colhead{T$_\mathrm{ex}$ (HC$_3$N)} & \colhead{T$_\mathrm{rot}$ (CH$_3$CCH)}  & \colhead{T$_\mathrm{d}$}  &  \colhead{T$_\mathrm{d}$(inner-outer) }\\
\colhead{(K)} &   \colhead{(K)} & \colhead{(K)} & \colhead{(K)} \\
\hline
11.5(1.0)  &  12.6(1.0)   &   13.8(0.2)  &   11.7-12.4\\
\hline
\end{tabular}
\\
\tablefoottext{ The T$_\mathrm{d}$(inner-outer) is model fitted from the SCUBA 850 $\mu$m map of L1489 \citep{2011ApJ...728..144F}.}\\
\end{table*}

\clearpage

\clearpage
      \begin{figure*}
        \centering
        \includegraphics[width=0.5\linewidth,height=1\linewidth]{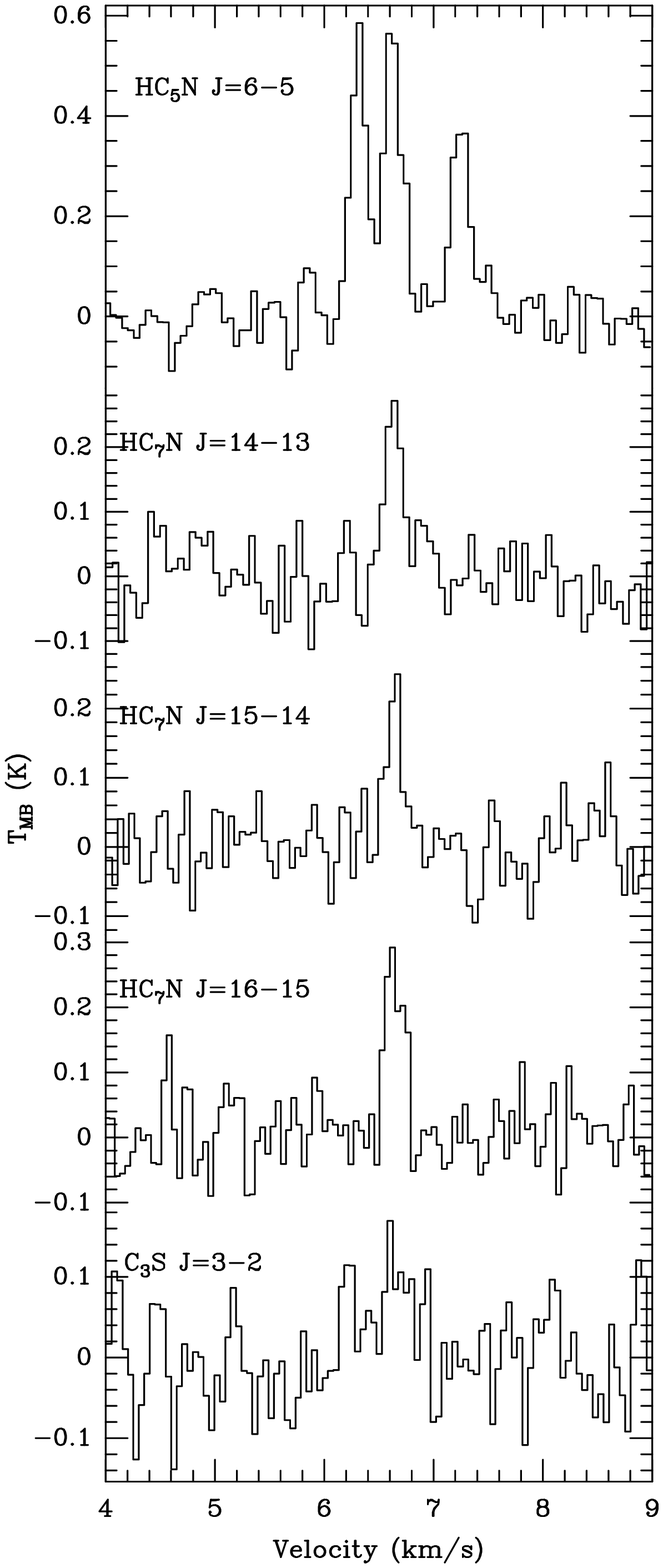}
        \hskip 0.2cm
        \includegraphics[width=0.3\linewidth,height=1\linewidth]{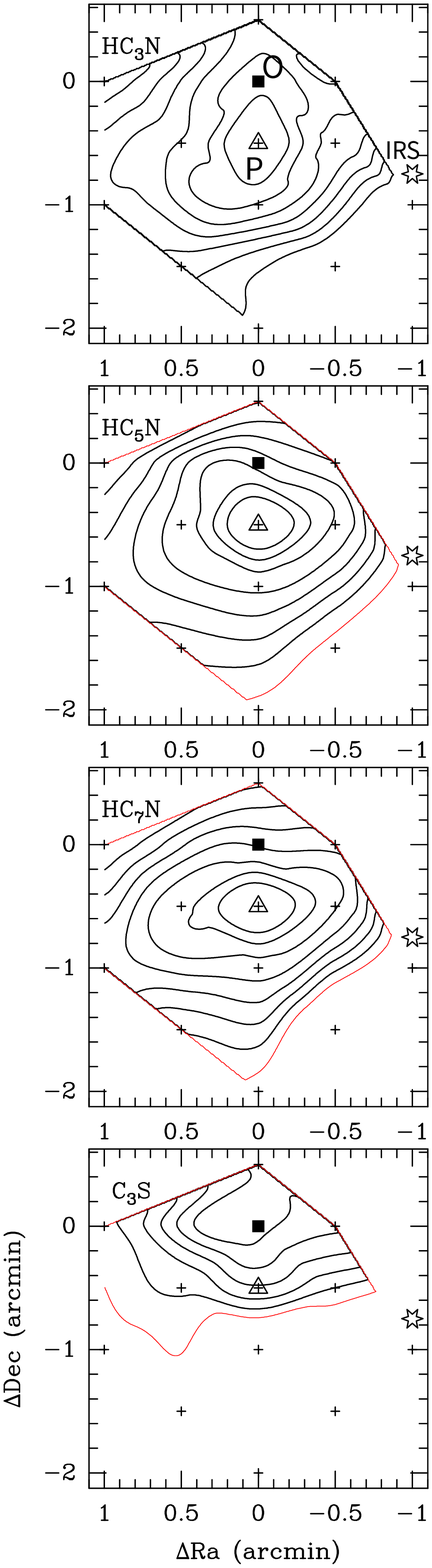}
        \caption{Emissions of \hctn~J=2-1, \hcfn~J=6-5, and \hcsn~J=14-13 as well as the \cts~J=3-2 in K$_u$ band.
     Left: Spectral lines at the P point. 
     The emission of \hctn~J=2-1, F=3-2 is shown in Fig. \ref{figure_wing}(a). Right: Intensity contours from 50$\%$ to 90$\%$ in steps of 10$\%$ of the peak value (see Table \ref{table_derived}) for \hctn~J=2-1, F=3-2, \hcfn~J=6-5, and \hcsn~J=14-13 as well as \cts~ J=3-2. Red contours represent 3 $\sigma$ of integrated emissions. The symbols are the same as  those in Fig. \ref{figure_wing}(c)
\label{figure_P_Ku}}.
\end{figure*}

\clearpage

      \begin{figure*}
         \centering
      \includegraphics[width=0.5\linewidth,height=1\linewidth]{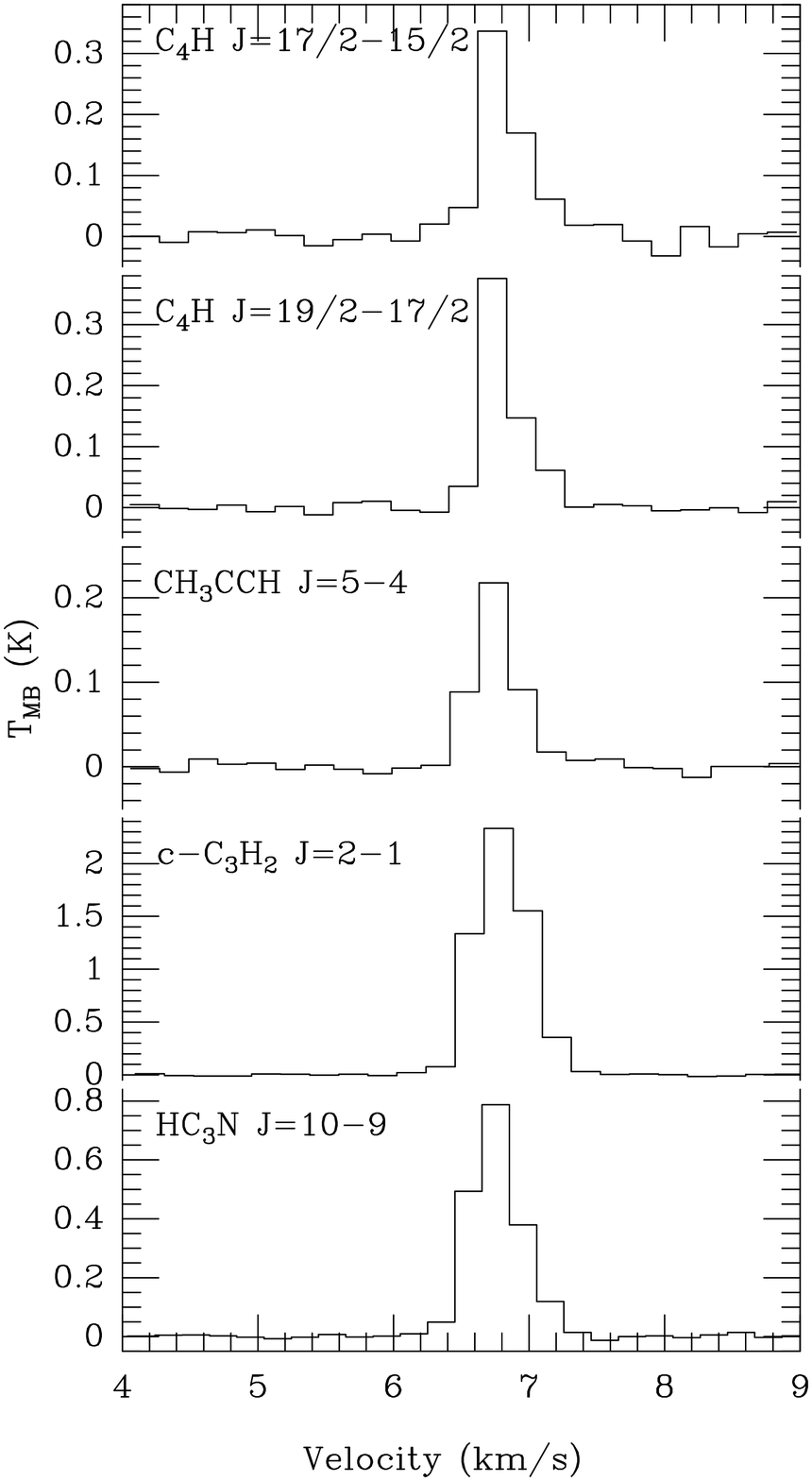}
       \hskip 0.2cm
        \includegraphics[width=0.3\linewidth,height=1\linewidth]{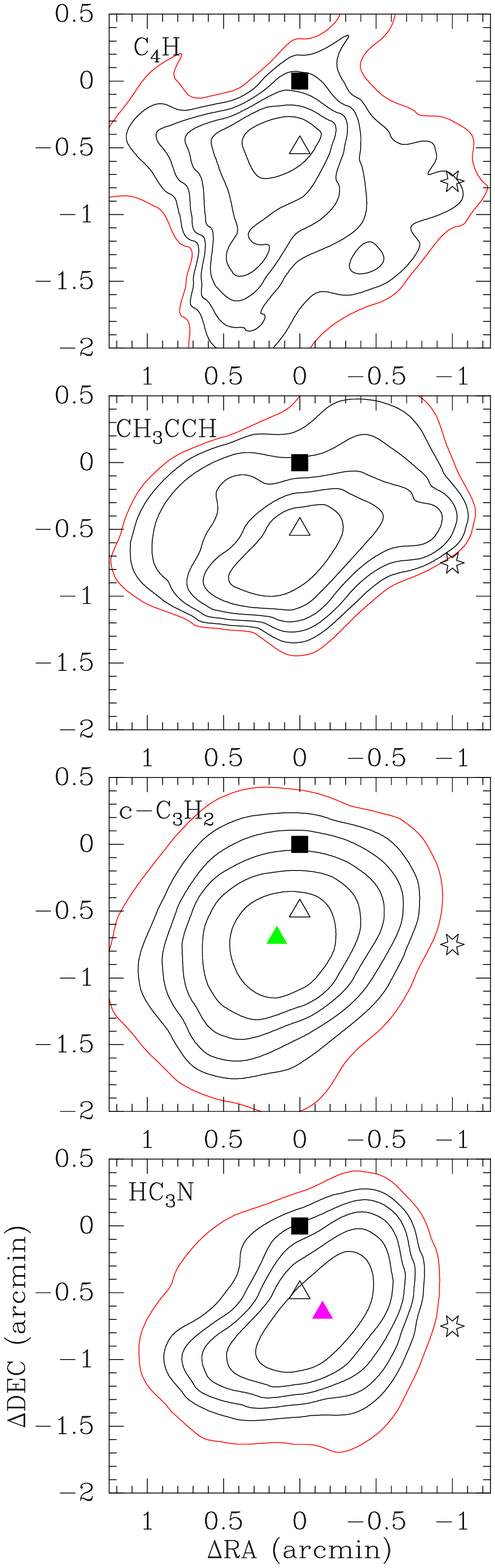}
        \caption{Emissions of the transitions in the 3 mm band. 
     Left: Spectral lines of the P point. Right: Emission intensity contours from
     50$\%$ to 90$\%$ in steps of 10$\%$ of the peak value (see Table \ref{table_derived}). \cfh~N=9-8, J=19/2-17/2 and \cfh~N=9-8, J=17/2-15/2 are summed up and shown
     as \cfh~intensity contours. Red contours represent 3 $\sigma$ of integrated emissions.  The symbols are the same as  those in Fig. \ref{figure_wing}(c). The green and pink filled triangles represent peak position of
       \cctht~J=2-1 and \hctn~J=10-9 emission, respectively.
       \label{figure_P_3mm} }

     \end{figure*}

\clearpage
     \begin{figure*}[htbp]
        \centering
        \setlength{\abovecaptionskip}{10pt}
        \setlength{\belowcaptionskip}{-18pt}
        \includegraphics[width=0.4\linewidth]{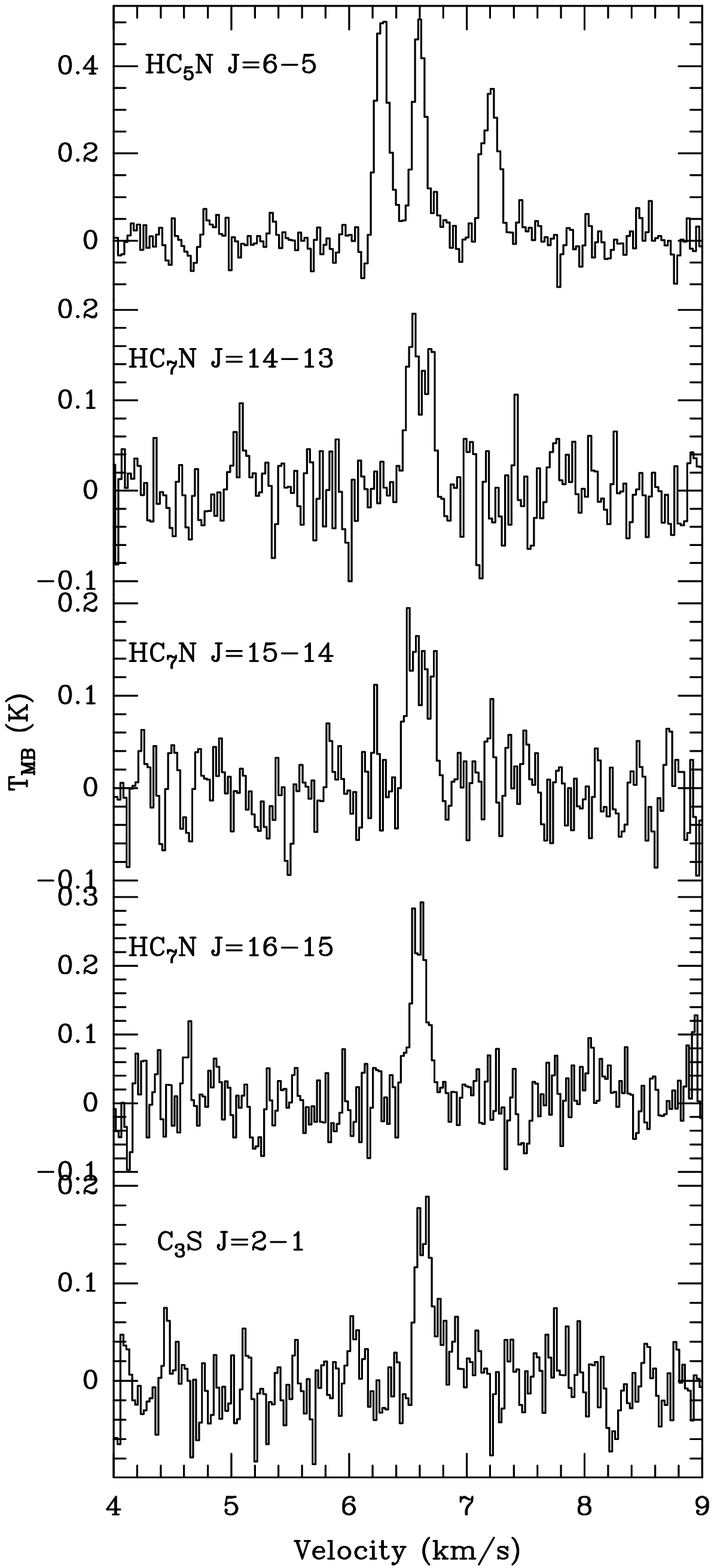}
       \includegraphics[width=0.4\linewidth]{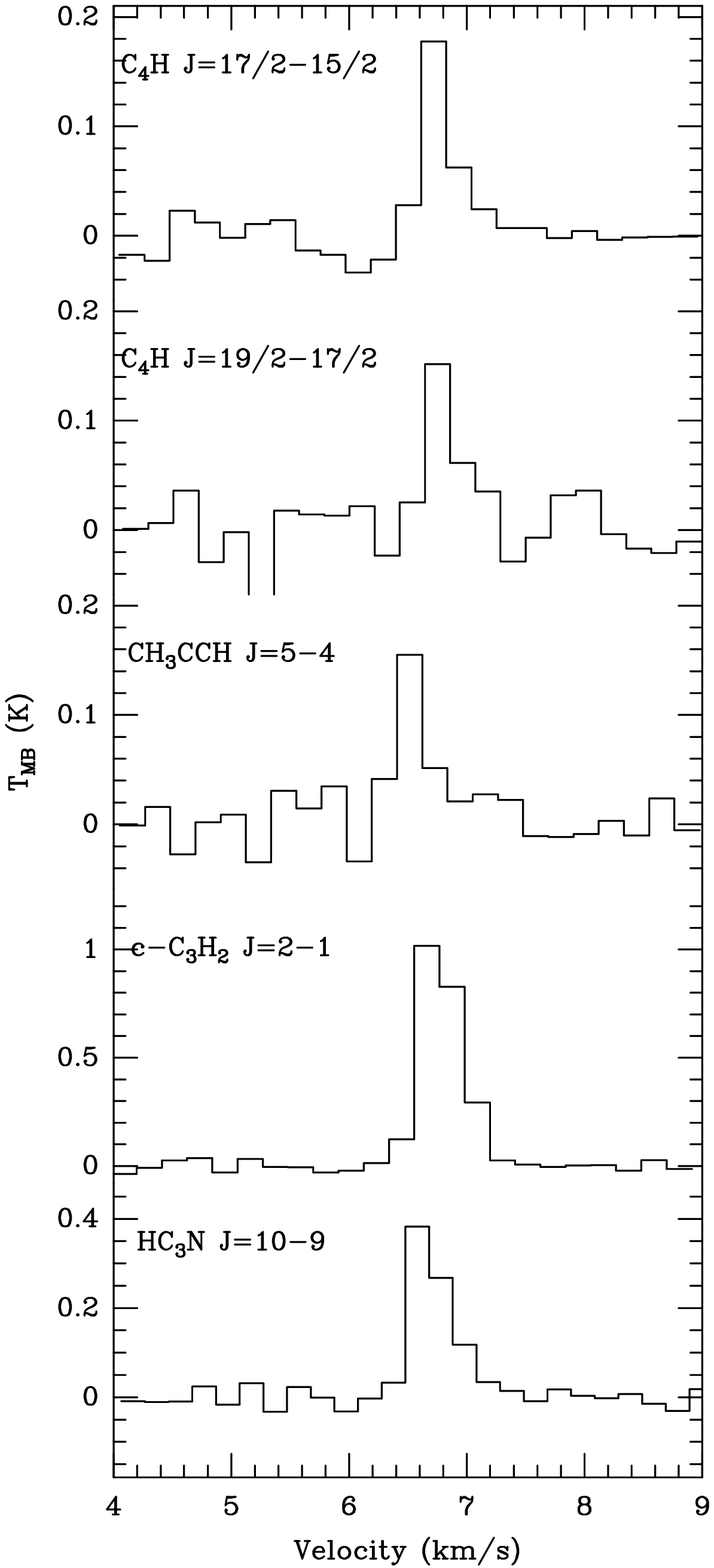}
        \caption{ Spectra at O position. Left: Spectra at K$_u$ band.  Right: Spectra at the 3 mm band detected with the 13.7 m telescope of PMO. \label{figure_O}}

     \end{figure*}

\clearpage
    \begin{figure*}
        \centering
     \includegraphics[width=0.35\linewidth]{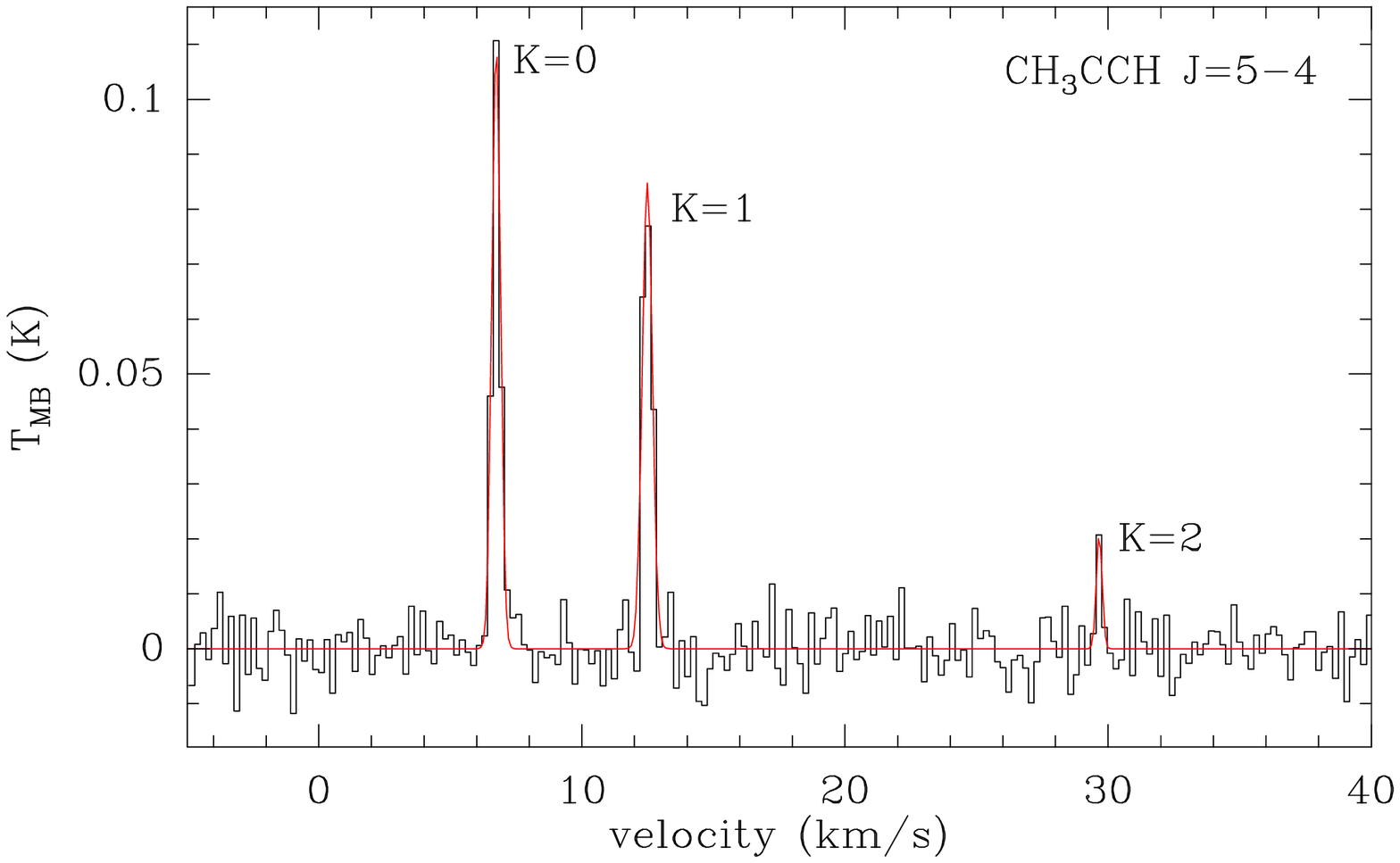}
     \includegraphics[width=0.26\linewidth]{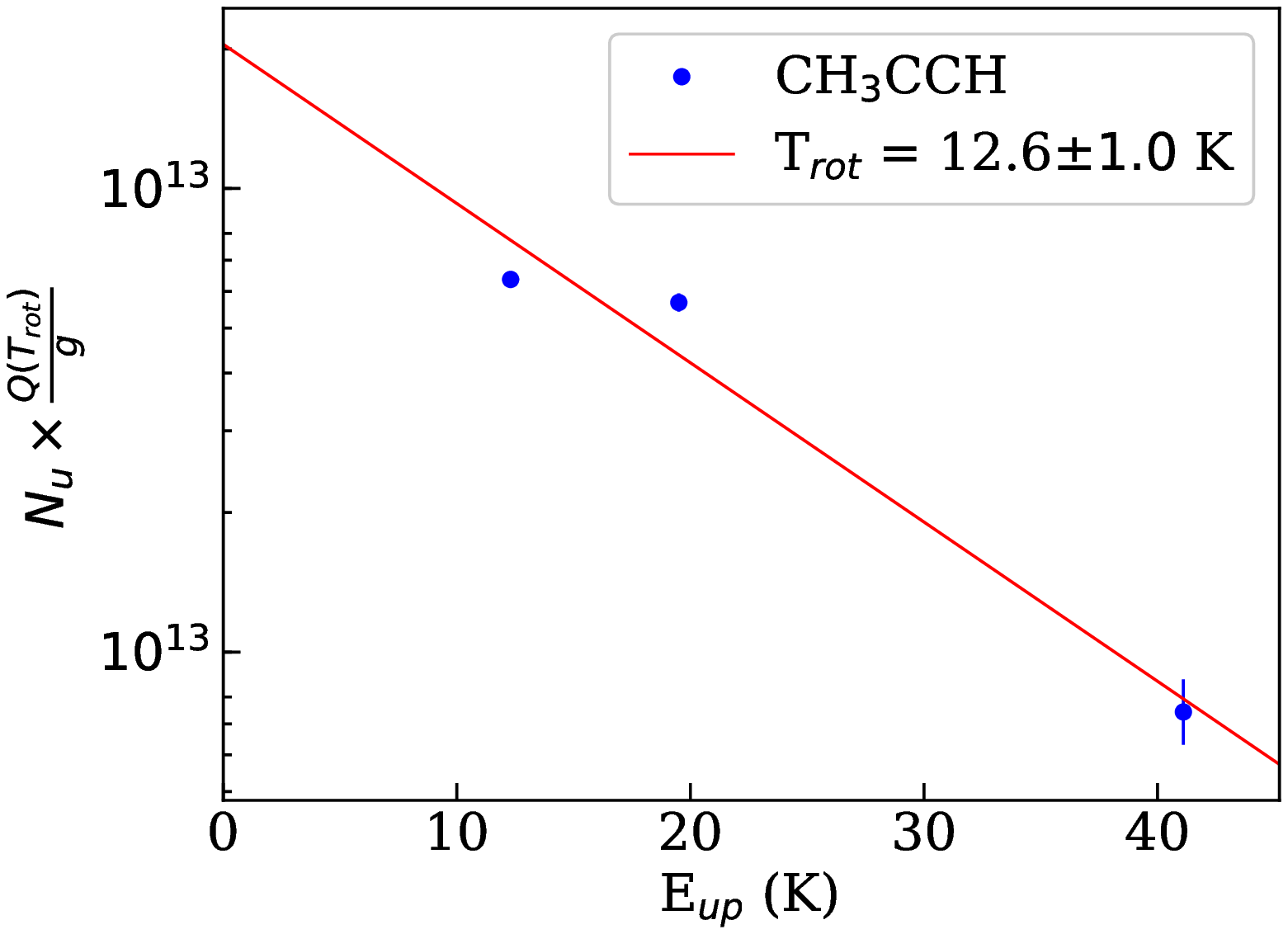}
     \includegraphics[width=0.27\linewidth]{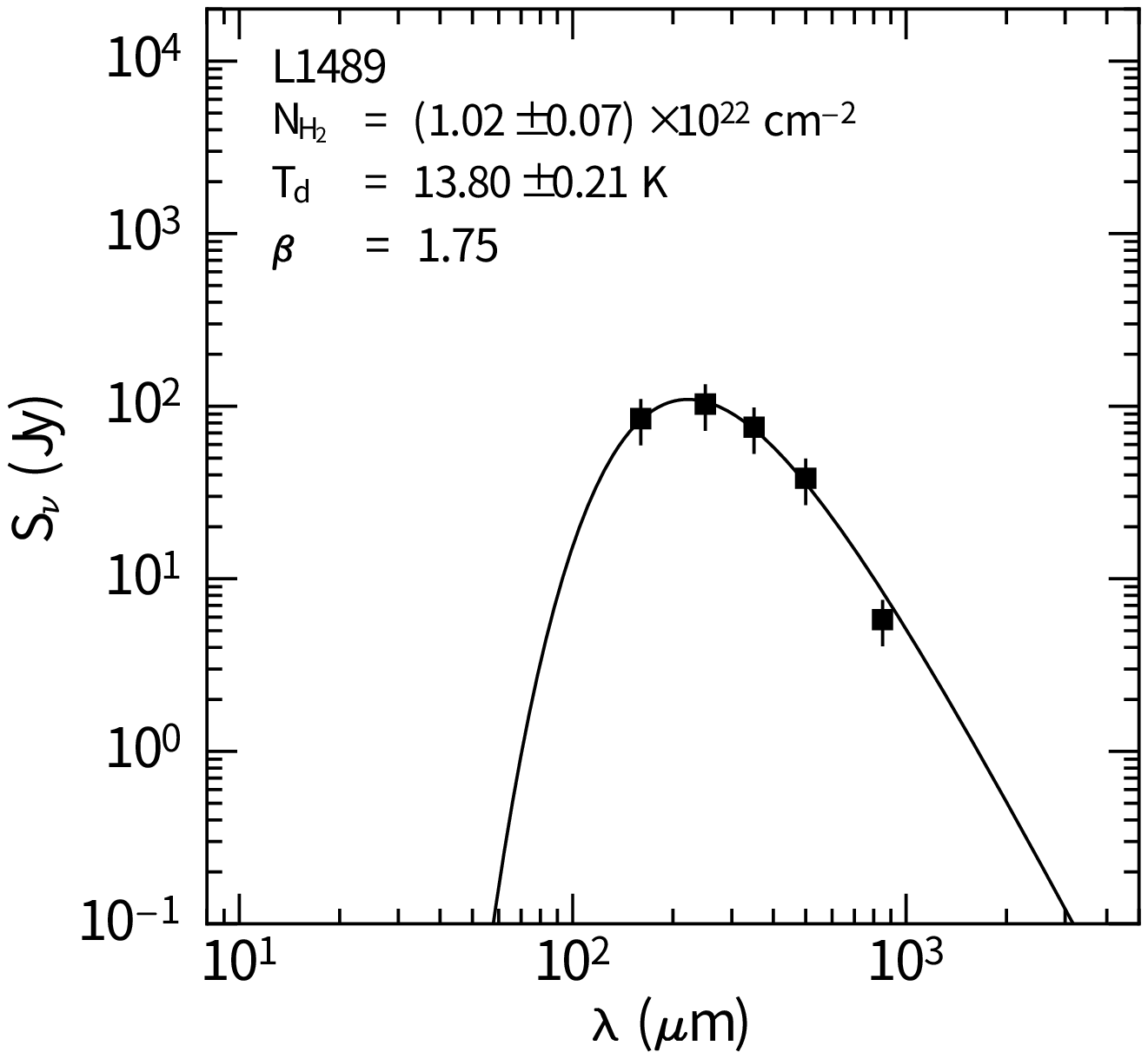}
        \caption{{ Left: Spectra of \chtcch~ J=5-4, K=0,1,2.
   Middle: Rotation temperature diagram of \chtcch~.
  Right: SED of the L1489 IRS from the PACS 160 $\mu$m and SPIRE wavelengths of Herschel as well as SCUBA 850 $\mu$m. The filled squares
represent the input fluxes. The line shows the best fitting of the gray-body model.}
   \label{figure_continuum}
   }
    \end{figure*}

\clearpage
\begin{figure*}
         \centering
       \includegraphics[width=0.48\linewidth,height=0.37\linewidth]{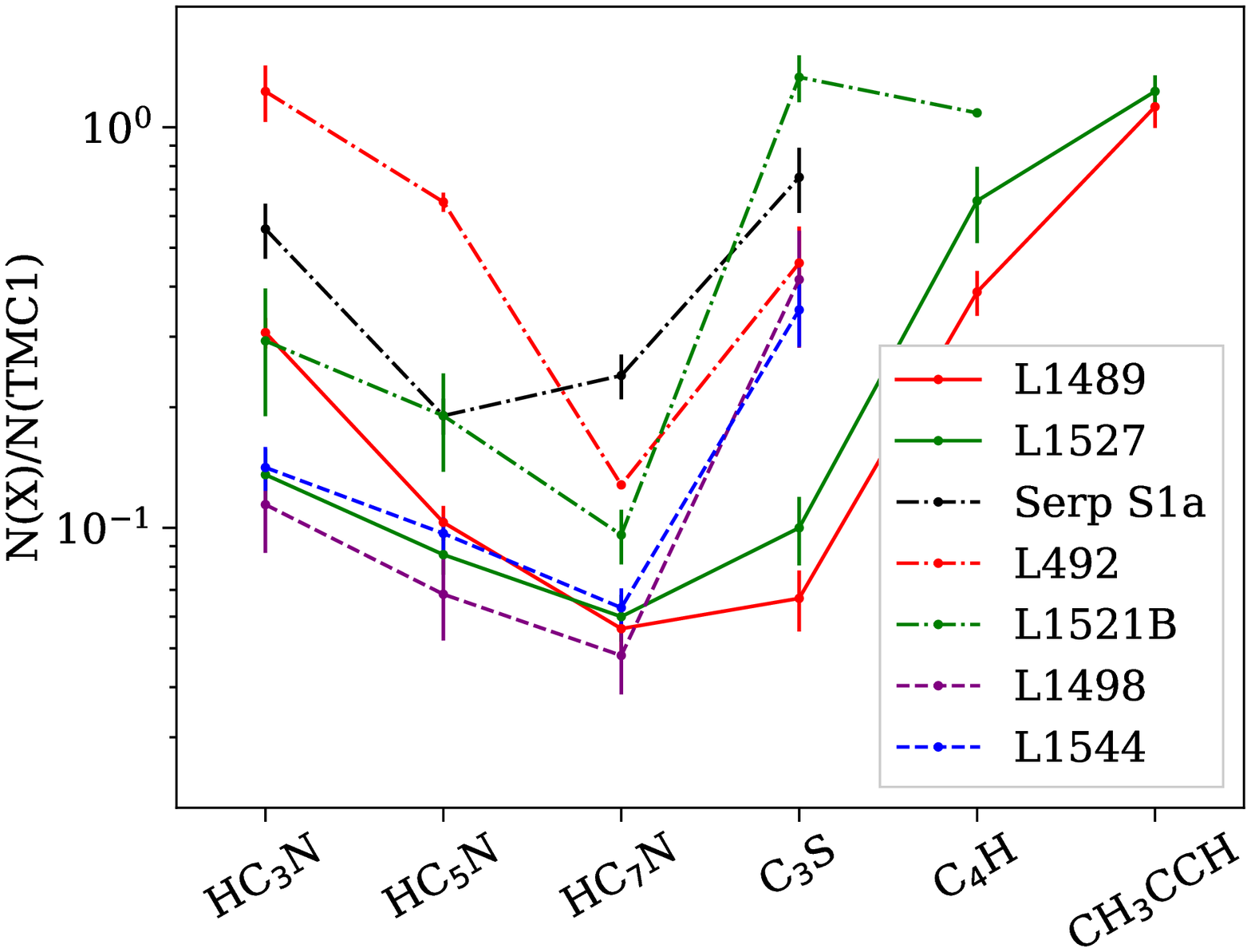}
        \caption{
     Comparison between the CCM column densities of L1489 EMC and typical CCM rich sources, normalized by the values of TMC-1 \citep{2004PASJ...56...69K,1992ApJ...392..551S}. The source names are denoted in the lower-right corner with different colors,
including starless cores Serpens South 1a \citep[Serp S1a;][]{2016ApJ...824..136L},
L492 \citep{2006ApJ...646..258H,2009ApJ...699..585H}, L1521B \citep{1992ApJ...392..551S,2004ApJ...617..399H}, L1498 \citep{1992ApJ...392..551S,1996ApJ...468..761K} and L1544 \citep{1992ApJ...392..551S} as well as WCCC source
L1527 \citep{2008ApJ...672..371S}.
The \hcsn~ column density of L1544 is derived from that of \hcfn~ assuming the ratio of HC$_{2n+1}$/HC$_{2n+3}$ (n=1,2) is constant \citep{1992ApJ...392..551S}.
The $N$(\cts) of L1527 is deduced from column densities of C$_2$S and the ratio of the C$_3$S/C$_2$S \citep{2008ApJ...672..371S,1992ApJ...392..551S}.
\label{figure_compare}}

     \end{figure*}

\clearpage
    \begin{figure*}
        \centering
     \includegraphics[width=0.3\linewidth]{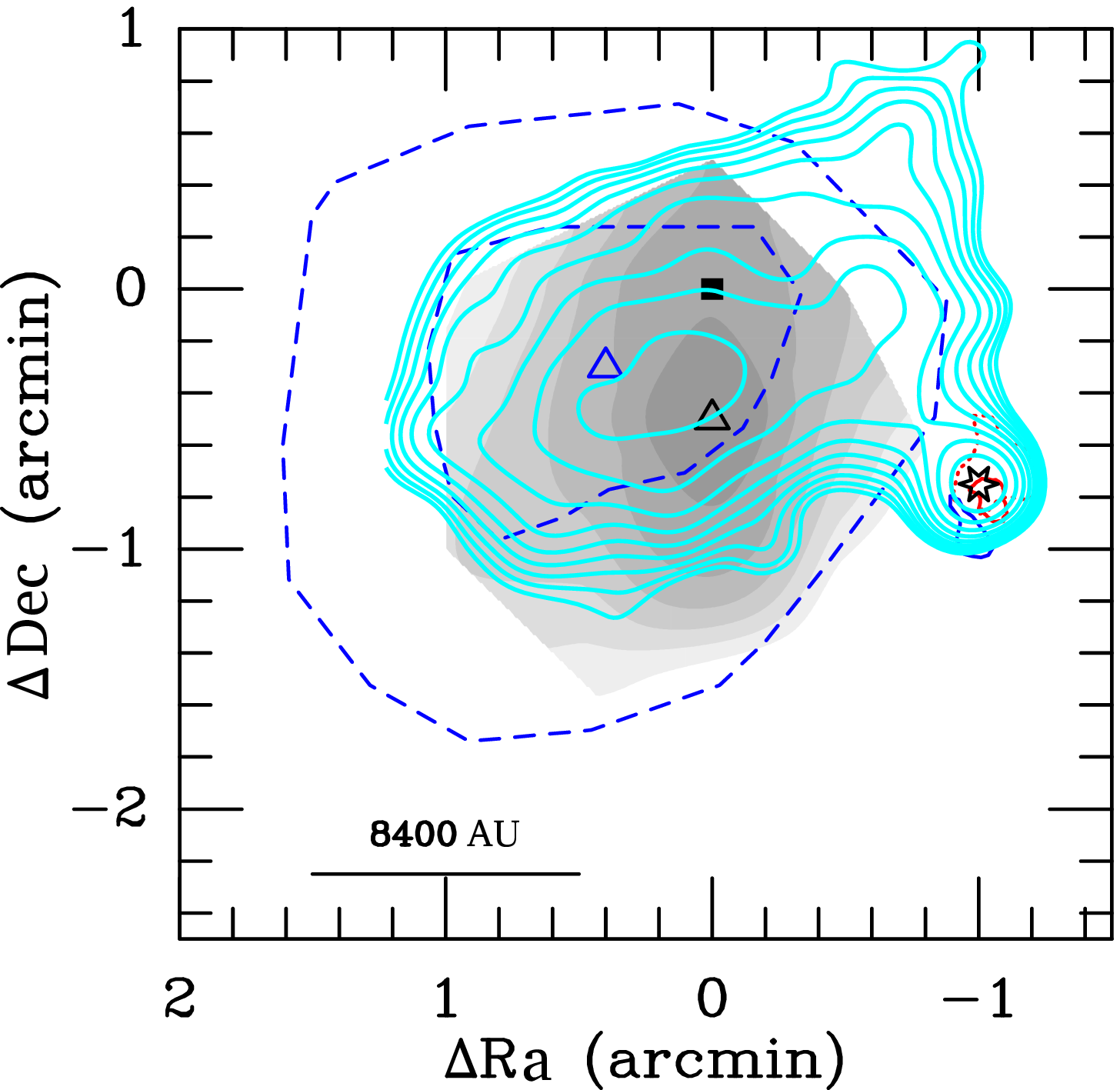}
     \includegraphics[width=0.3\linewidth]{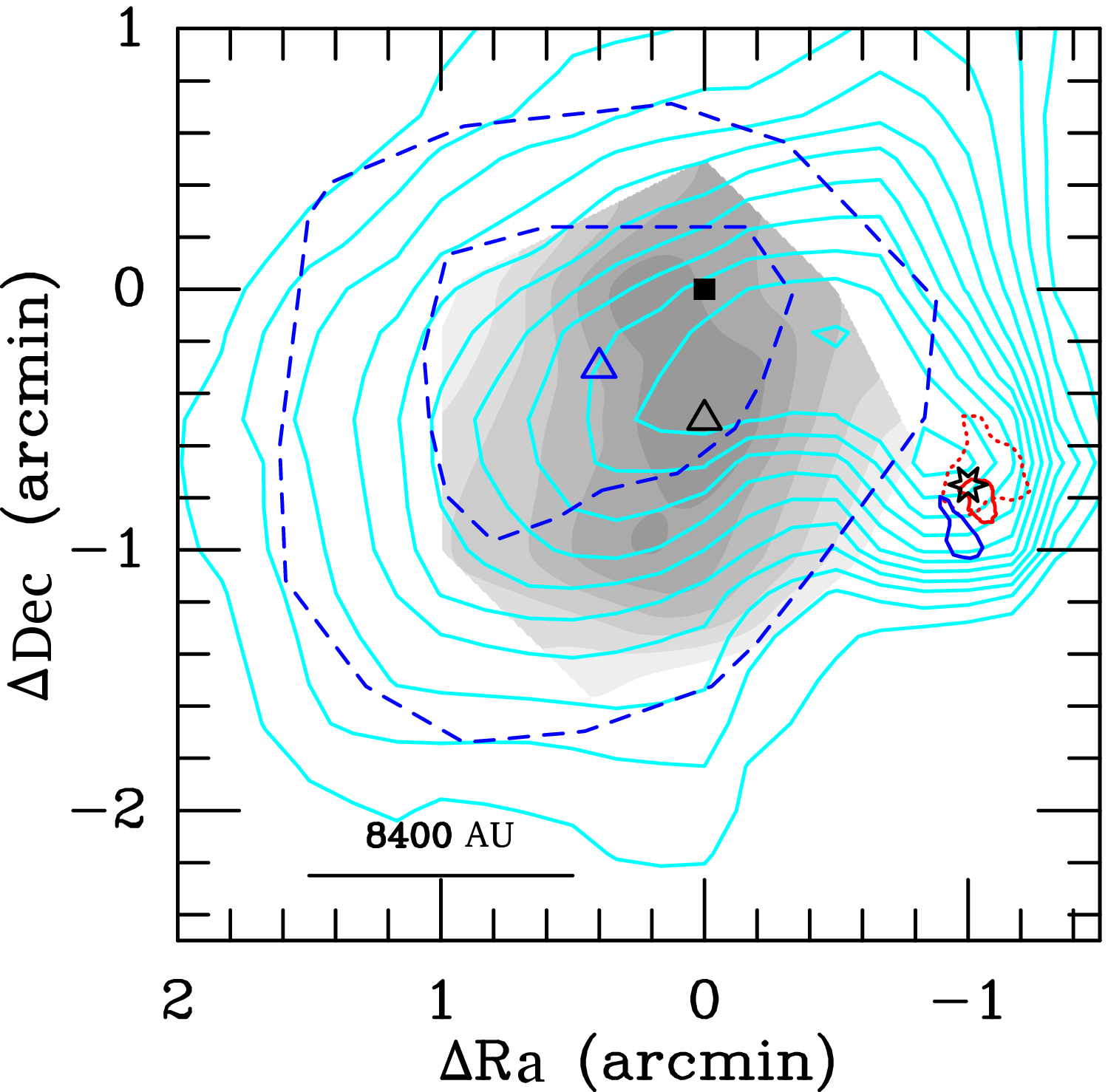}
     \includegraphics[width=0.3\linewidth]{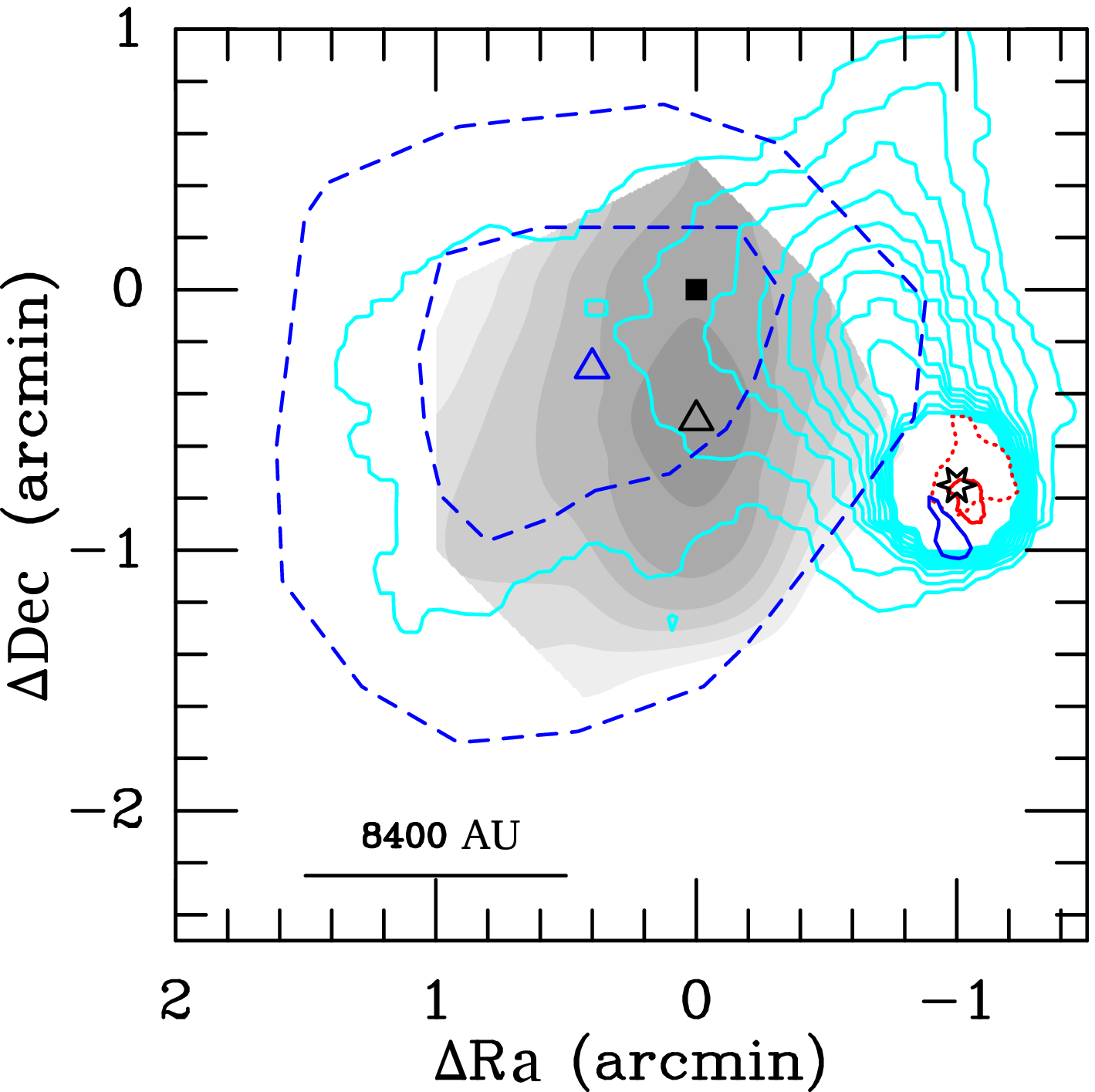}
        \caption{Left: The  \hctn~core (gray-scale) and the NH$_3$ (1,1) core (blue dashed lines, quoted from \cite{1988ApJ...324..907M} ) are overlaid on the JCMT SCUBA 850 $\mu$m continuum data { from JCMT proposal ID M97AN16} \citep{2000ApJ...534..880H} (cyan contours, { evenly stepped from 0.075 to 0.75 Jy/beam in log-scale}). The gray and blue triangles denote the peaks of the EMC (P point) and NH$_3$, respectively. The IRS is also marked on the map (hexagonal star). The blue and red wings of the CO (3-2) outflow (blue solid and red dotted lines, quoted from \cite{1998ApJ...502..315H}) and the medium infalling lobe (red solid line, quoted from \cite{2014ApJ...793..1Y}) were also overlaid on the figure. Middle: As in left panel except cyan contours representing Herschel SPIRE 500 $\mu$m  continuum map, from 1.5 Jy/beam to 5 Jy/beam stepped by 0.3 Jy/beam. Right: As in left panel except cyan contours representing Herschel SPIRE 250 $\mu$m  continuum contours, from 1.5 Jy/beam to 5 Jy/beam stepped by 0.3 Jy/beam. Fluxes around the IRS are centrally peaked but higher level contours are not shown. 
   \label{figure_continuum}
   }
    \end{figure*}

\end{document}